\documentclass[prd,twocolumn,preprintnumbers,superscriptaddress,%
floatfix,showpacs,showkeys,nofootinbib]{revtex4}

\usepackage[figuresright]{rotating}
\usepackage{amssymb,amsmath,amsfonts,amsthm,graphicx,psfrag}

\newcommand{\beq}{\begin{equation}}
\newcommand{\eeq}{\end{equation}}
\newcommand{\bea}{\begin{eqnarray}}
\newcommand{\eea}{\end{eqnarray}}
\newcommand{\Fig}[1]{Fig.~\ref{#1}}
\newcommand{\Tab}[1]{Table~\ref{#1}}
\newcommand{\Sec}[1]{Section~\ref{#1}}
\newcommand{\Eq}[1]{Eq.~(\ref{#1})}

\newcommand{\fc}{\mathbf{fc}}
\newcommand{\bc}{\mathbf{bc}}

\newcommand{\re}{\operatorname{\mathfrak{Re}}}
\newcommand{\tr}{\operatorname{tr}}

\def\Pi{\mathcal{P}_\infty}

\usepackage{color}
\begin{document}

\preprint{BI-TP 2010/25, HU-EP-10/51, LU-ITP 2010/004}

\title{$SU(3)$ Landau gauge gluon and ghost propagators \\ 
     using the logarithmic lattice gluon field definition}

\author{Ernst-Michael~Ilgenfritz}
\affiliation{Universit\"at Bielefeld, Fakult\"at f\"ur Physik, 
  33615 Bielefeld, Germany}
\affiliation{Humboldt-Universit\"at zu Berlin, Institut f\"ur Physik, 
  12489 Berlin, Germany}

\author{Christoph~Menz}
\affiliation{Humboldt-Universit\"at zu Berlin, Institut f\"ur Physik, 
  12489 Berlin, Germany}
\affiliation{Potsdam Institut f\"ur Klimafolgenforschung, 
  14473 Potsdam, Germany}

\author{Michael~M\"uller-Preussker}
\affiliation{Humboldt-Universit\"at zu Berlin, Institut f\"ur Physik,
  12489 Berlin, Germany}

\author{Arwed~Schiller}
\affiliation{Universit\"at Leipzig, Institut f\"ur Theoretische Physik, 
  04009 Leipzig, Germany}

\author{Andr\'e~Sternbeck}
\affiliation{Institut f\"ur Theoretische Physik, Universit\"at Regensburg, 
  93040 Regensburg, Germany}

\date{February 28, 2011}

\begin{abstract}
We study the Landau gauge gluon and ghost propagators of $SU(3)$ gauge
theory, employing the logarithmic definition for the lattice gluon
fields and implementing the corresponding form of the Faddeev-Popov matrix. 
This is necessary
in order to consistently compare lattice data for the bare propagators
with that of higher-loop numerical stochastic perturbation theory
(NSPT). In this paper we provide such a comparison, and introduce what
is needed for an efficient lattice study. When comparing our data for
the logarithmic definition to that of the standard lattice Landau
gauge we clearly see the propagators to be multiplicatively
related. The data of the associated ghost-gluon coupling matches up
almost completely. For the explored lattice spacings and sizes
discretization artifacts, finite-size and Gribov-copy effects are
small. 
At weak coupling and large momentum, the bare propagators
and the ghost-gluon coupling are seen to be approached by 
those of higher-order NSPT.
\end{abstract}



\maketitle


\section{Introduction}
\label{sect:Introduction}

During the last years lattice studies on the Landau gauge gluon and
ghost propagators have clearly
\cite{Cucchieri:2007rg,Cucchieri:2008fc,Bogolubsky:2009dc,Bornyakov:2009ug}
not confirmed the asymptotic infrared behavior as postulated by the
Gribov-Zwanziger scenario \cite{Gribov:1977wm,Zwanziger:1993dh} or
required by the Kugo-Ojima confinement criterion
\cite{Kugo:1979gm}. The inconsistency seems to be related to the
treatment of the Gribov problem, i.e., the non-uniqueness of the
Landau gauge condition in the non-perturbative regime (at least in
finite volumes). In our opinion, it is still unclear,  
whether the lattice formalism---for a fixed gauge---requires a reformulation 
in order to become consistent with a BRST-invariant continuum theory or 
whether the continuum approach itself needs to be reformulated 
(see also~\cite{Fischer:2007pf,vonSmekal:2008ws}). 
We would like to stress that until now no complete gauge-fixing prescription
for Landau gauge has been agreed upon (see, e.g.,
\cite{Maas:2009se,Maas:2009ph} and references therein). At the same
time it is clear that the gauge-invariant  
lattice formulation of $SU(N)$ Yang-Mills theories {\it does provide 
confinement} of quarks {\it and} gluons. 
Also in the context of gauge-variant approaches, there are clear signals that
quarks and gluons (and ghosts) are not part of the physical spectrum, 
for example, by the violation of reflection positivity
~\cite{Cucchieri:2004mf,Sternbeck:2006cg,Bowman:2007du}.
The exact mechanism of confinement, however, is still unknown.
It should be added though that a direct link between quark confinement,
as measured by the Polyakov-loop order parameter, and the infrared behavior
of ghost and gluon Green's functions has been established 
recently~\cite{Braun:2007bx}.

However, the infrared asymptotics can only be one aspect. Another, by
no means less important, should be the computation of QCD's elementary
two- and three point functions in the intermediate (around $1
\mathrm{~GeV}$) and ultraviolet momentum region. This would allow for
example to determine essential phenomenological parameters, like the
QCD scale $\Lambda_{\overline{\mathsf{MS}}}$, or gluon and quark
condensates (see, e.g., 
\cite{Becirevic:1999hj,Boucaud:2005rm,Boucaud:2005gg,Boucaud:2008gn,Sternbeck:2010xu,Blossier:2010ky}).
Moreover, such calculations are important to arrive at renormalized
propagators, and eventually also at vertex functions, which,
calculated on the lattice and extrapolated to the continuum limit,
could serve as input to a Bethe-Salpeter or Faddeev equations based
hadron phenomenology (see, e.g., \cite{Alkofer:2009jk} for a status
report).

Therefore, the validity of multiplicative renormalizability in the
nonperturbative regime as well as the speed of convergence to the
continuum limit, and its uniqueness, are essential questions that need
to be addressed on the lattice in this context. For the gluon field
this has been done in the
past~\cite{Giusti:1998ur,Petrarca:1998je,Furui:1998cg,Cucchieri:1999dt,
Bogolubsky:2002ui}, 
in particular there with respect to the question of universality of its
definition on the lattice. Also, more recently, first steps towards
continuum-limit-extrapolated lattice data for the gluon and ghost
propagators have been presented \cite{Bornyakov:2009ug,Bogolubsky:2009qb}. 

With our study we intend to provide further input to such projects,
placing here particular emphasis on connecting lattice Monte Carlo
studies with those of lattice perturbation theory (LPT). Specifically
we will confront its numerical variant, the numerical stochastic
perturbation theory (NSPT) \cite{DiRenzo:2009ni,DiRenzo:2010cs}, with
lattice Monte Carlo (MC) data of the gluon and ghost propagators and
the associated ghost-gluon coupling in Landau gauge
\cite{vonSmekal:1997is}. This will help to quantify the region where
predictions of NSPT are valid. 

So far, most MC studies of these gauge-variant objects have used (what 
we call below) the \emph{linear definition}
\beq
A^{\rm (lin)}_{x+\frac{\hat{\mu}}{2},\mu}  =  \frac{1}{2iag_0} 
\left( U_{x,\mu} - U^{\dagger}_{x,\mu} \right)\Big|_{\rm traceless}
\label{eq:Alin}
\eeq
with $a$, $g_0$ denoting the lattice spacing and the bare coupling, 
respectively. On the lattice, however, the definition of the gluon
field (and also of the Faddeev-Popov operator) is in no way
unique. One can of course equally well use another definition, for
example, that of the modified lattice Landau gauge 
\cite{vonSmekal:2007ns},%
\footnote{Note that the modified lattice Landau gauge (MLG) actually
  provides more than just another lattice discretization of the involved fields
  \cite{vonSmekal:2007ns}. It is a potential 
  candidate to overcome the 0/0 Neuberger problem on the
  lattice~\cite{Neuberger:1986vv,Neuberger:1986xz}. Nonetheless, one
  could borrow the MLG lattice definitions for the gluon field and
  Faddeev-Popov operator if an alternative discretization of
  the standard lattice Landau gauge is desired. So far this would be
  possible for a $U(1)$ or $SU(2)$ gauge theory, and has been done for
  the latter for example in \cite{Sternbeck:2008mv}).} 
or the
\emph{quadratic definition}
\beq
A^{\rm (quad)}_{x+\frac{\hat{\mu}}{2},\mu}  =  \frac{1}{4iag_0} 
\left[ (U_{x,\mu})^2 - (U^{\dagger}_{x,\mu})^2 \right]\Big|_{\rm traceless}
\label{eq:Aquad}
\eeq
which has been studied and compared with the linear definition in
Refs.~\cite{Giusti:1998ur,Petrarca:1998je}. It was also shown there
both converge towards the same continuum limit.

Yet another definition is known as the \emph{logarithmic definition}
of the lattice gluon field,
\beq
A^{\rm (log)}_{x+\frac{\hat{\mu}}{2},\mu} =
\frac{1}{iag_0} \log \left( U_{x,\mu} \right) \, .
\label{eq:Alog}
\eeq
This definition has been put forward for lattice MC studies by Furui
and Nakajima (see, e.g.,
\cite{Furui:1998cg,Nakajima:2000nr,Furui:2003jr,Furui:2004cx,
Furui:2005bu}). It is also the definition being
used in lattice perturbation theory (see, e.g., the
monograph~\cite{Rothe:1997kp}) and NSPT. Since we are aiming at a
quantitative comparison of NSPT with lattice MC data,
below we will mainly concentrate on this definition.

The respective control functional for the Landau gauge, the quadratic norm
\beq
||\Delta||^2 = \sum_x~\tr \Delta(x) \Delta^{\dagger}(x) \, ,
\label{eq:deltanorm}
\eeq
is given in terms of the lattice divergence
\beq
\Delta = 
\left( \sum_\mu \partial_\mu A_\mu \right)(x) \equiv 
\sum_\mu \left( A_{x+\frac{\hat{\mu}}{2},\mu} -
                A_{x-\frac{\hat{\mu}}{2},\mu} \right) \ne 0 \, ,
\label{eq:divgeneral}
\eeq
for each of these definitions. 

When expanded in terms of the lattice spacing all these definitions of
the gluon field agree at leading order, but differ beyond that (see,
e.g., \cite{Sternbeck:2008mv}). It is because of these differences why
the corresponding Jacobian factors (Faddeev-Popov determinants) differ
in the integration measure and why the respective lattice Landau gauge
condition ($||\Delta||^2$) cannot be simultaneously satisfied (see, e.g.,
\cite{Giusti:1998ur,Petrarca:1998je} or below). For a consistent setup
of Landau gauge on the lattice one therefore has to employ the
corresponding lattice expressions for the gauge functional and the
Faddeev-Popov operator. These are available in the literature for all
above-mentioned approaches. For practical reasons, though, mostly the
linear definition has been adopted.

However, in order to assess the genuine non-perturbative effects in
the measured two- and three-point functions, it is desirable to have
as a reference point an understanding of the perturbative behavior of
these functions in higher-order lattice perturbation theory.  In
recent years such higher-loop results for the lattice gluon and ghost
propagators became available using numerical stochastic perturbation
theory (NSPT) \cite{DiRenzo:2009ni,DiRenzo:2010cs}. These results are
for individual momenta at a fixed lattice size and are usually
obtained for the logarithmic definition.  The advantage of NSPT is
that the loop-order that can be achieved only depends on the available
computational resources. There are no other restrictions. However, if
one wants to confront the unrenormalized NSPT results directly with
corresponding data from lattice Monte-Carlo (MC) simulations, one
cannot expect the convergence of the former to the bare MC data, when
different lattice gluon field definitions are employed. Of course, if
multiplicative renormalization is in place the bare propagators for
the different definitions are related through finite
renormalizations, but for a comparison with NSPT this would bring
additional uncertainties. For a quantitative comparison it is thus
desirable to use the same lattice definition of the respective gluon
field and the Faddeev-Popov operator. In addition, lattice NSPT results 
take the influence of the hypercubic group into account and thus allow 
for a comparison of the respective propagators at more off-diagonal 
momenta which are usually excluded by cuts.

We therefore find it worth to
complement the existing data for the Landau gauge gluon and ghost
propagators by new sets for the logarithmic definition of the lattice
gluon fields and the corresponding Faddeev-Popov operator. The
bare propagators can then be confronted directly (i.e., without any
additional renormalization) to the results from NSPT. We think that
this way much more precise information can be obtained on the regime
where NSPT holds. 

Note that for a renormalization-group invariant, like for example for
the ghost-gluon coupling, denoted 
$\alpha^{\mathsf{MM}}_s(q^2)$ in the minimal MOM scheme
\cite{vonSmekal:2009ae, Sternbeck:2010xu}, no rescaling of the data
would be necessary, when comparing data obtained for
different lattice discretization.\footnote{It is important though, that 
the bare lattice gluon and ghost dressing functions
is for the same type of lattice discretization, and the correct
tree-level value is obtained when the links are set to
one~(see also \cite{Sternbeck:2008mv}).} Data for this
coupling, extracted either 
for the linear or the logarithmic definition, should match up
completely, apart from artifacts due to the lattice
discretization. This has been argued and, for $SU(2)$, also explicitly
shown already in \cite{Sternbeck:2008mv}. This coupling thus serves a
reasonable object to systematically investigate discretization
effects. Below we will further corroborate these findings for the
gauge group $SU(3)$ and compare the available data sets to
corresponding ones of NSPT.

\medskip

Altogether, our aim is threefold: 
\begin{enumerate}
\item To check universality of the employed lattice definitions
  by comparing MC results obtained for the logarithmic and linear
  definition. Such a study is not really new, but we do it
  simultaneously for the gluon \emph{and} ghost propagator and the
  associated ghost-gluon coupling. We confirm that, at least in the
  momentum ranges considered throughout this paper, the propagators are
  indeed related to each other by multiplicative renormalization, and
  that these renormalization constants are such that they exactly
  cancel each other in the coupling as expected.

\item To compare nonperturbative and perturbative lattice results
  obtained for exactly the same set of parameters. We emphasize that (in
  this study) we are not intending to fit lattice results in the
  perturbative range with higher-order perturbation theory as obtained
  in the continuum by \cite{Chetyrkin:2000dq,Chetyrkin:2004mf,vonSmekal:2009ae}. 
  This has been the intention in the lattice references  
  \cite{Furui:2003jr,Becirevic:1999hj,Boucaud:2005rm,Boucaud:2005gg,Boucaud:2008gn,Sternbeck:2010xu,Blossier:2010ky}.
  Here we are rather interested to locate the momentum region where predictions of NSPT 
  are still a good approximation to Monte Carlo results. 
\item To report in some detail on the realization and performance
  of the gauge-fixing algorithm required for the logarithmic
  definition, as well as to assess the importance of the Gribov
  ambiguity in this setting.
\end{enumerate}

The paper is organized as follows: In \Sec{sec:implementation} we briefly
review lattice Landau gauge for the linear and the logarithmic definition  
of gluon fields. \Sec{sec:algorithms} is then devoted to
gauge-fixing algorithms, first for $A^{\rm (lin)}_{x+\hat{\mu}/2,\mu}$, 
and then for $A^{\mathrm{(log)}}_{x+\hat{\mu}/2,\mu}$. 
For the latter we test three
different implementations: an unaccelerated, a Fourier-accelerated
and a multigrid-accelerated gauge-fixing algorithm. We will argue that
one should preconditioning these by first fixing the gauge field
configuration such that $A^{\mathrm{(lin)}}_{x+\hat{\mu}/2,\mu}$
is transversal, before the actual gauge-fixing of
$A^{\mathrm{(log)}}_{x+\hat{\mu}/2,\mu}$ starts. 
A comparison of the three different
implementations including their parameters is given
at the end of \Sec{sec:algorithms}. In
\Sec{sec:compareobservabledefinitions} we present expressions for the
gluon and ghost propagators for the logarithmic definition. A
brief study on the impact of the Gribov ambiguity is also discussed there.
Further results are then presented in \Sec{sec:MCresults}, where we
first compare the gluon and ghost propagators for the logarithmic
definition with those for the linear definition, and then
discuss lattice discretization and finite-volume effects. We will
demonstrate that data for the coupling $\alpha^{\mathsf{MM}}_s(q^2)$ 
matches up for 
both definitions without any rescaling. In
\Sec{sec:compareresultslogversusNSPT} we 
finally confront our MC results for the propagators and the coupling
to the recent results from NSPT
\cite{DiRenzo:2009ni,DiRenzo:2010cs}. We will argue that for large 
$\beta=6/g_0^2$ and large momenta the MC data must be restricted to
the trivial (i.e.\ real-valued) Polyakov loop sector in order 
to reach good agreement.\footnote{Note this was argued for $SU(2)$ already
in \cite{Damm:1998pd}.} In \Sec{sec:conclusions} we will draw our  
conclusions. A detailed discussion of the multigrid-accelerated gradient
algorithm follows in the Appendix.

\section{Lattice implementations of Landau gauge}
\label{sec:implementation}

On the lattice the fundamental degrees of freedom are the link
variables $U_{x,\mu}$, which are elements of the $SU(3)$ gauge
group. If one is interested in gauge-variant quantities, like for
example the gluon propagator, one has to adopt a definition for the
gluon field $A_{x+\hat{\mu}/2,\mu}$ in terms of these links. 
Above, we mentioned
three ways of defining a gluon field on the lattice, the linear and
logarithmic definition will be considered below.

\subsection{Linear definition of the gluon field}

We first recall the linear definition [\Eq{eq:Alin}].
For this the Landau gauge condition on the lattice (setting the lattice spacing
$a=1$),
\bea
\Delta^{\rm (lin)}(x) & = & 
\left( \sum_\mu \partial_\mu A^{\rm (lin)}_\mu \right)(x) \nonumber \\
& \equiv & 
  \sum_\mu \left( A^{\rm (lin)}_{x+\frac{\hat{\mu}}{2},\mu} -
                  A^{\rm (lin)}_{x-\frac{\hat{\mu}}{2},\mu} \right) = 0 \, ,
  \label{eq:divAlin}
\eea
is realized, if the gauge functional 
\beq
F^{\rm (lin)}_U[g] = \frac{1}{4V}
\sum_{x,\mu}~\left(1 - \frac{1}{3}~\re \tr~{}^gU_{x,\mu}\right) 
\label{eq:Flin}
\eeq
is in a (local) minimum for a given gauge-fixed configuration
\beq
{}^gU^{\hphantom{\dagger}}_{x,\mu} = 
g^{\hphantom{\dagger}}_x U^{\hphantom{\dagger}}_{x,\mu} 
g^{\dagger}_{x+\hat{\mu}} \,.
\label{eq:gtrafo}
\eeq
Here, $g_x \in SU(3)$ is the gauge transformation field that -
finally - should put the unfixed gauge field $U_{x,\mu}$ to Landau 
gauge. As there are many local minima (``Gribov copies'') 
we assume that unique gauge fixing is achieved by searching
for the global minimum for each configuration $U$.
In practice, one can only try to get as close as possible 
to the global extremum. We call this general prescription 
``minimal Landau gauge''. The minimization of $F^{\rm (lin)}_U[g]$ 
can be accomplished by an overrelaxation algorithm or a combination of a 
simulated annealing and overrelaxation (see below).
This allows to fulfil the Landau gauge condition (\ref{eq:divAlin}) 
with the required local numerical precision.

\subsection{Logarithmic definition of the gluon field}

In the continuum, Landau gauge can also be formulated as a 
minimization problem of the functional
\beq
F^{\rm (cont)}_A[g] = \frac{1}{N_c} 
\sum_{\mu}~\int d^4x~\tr \left[ {}^g \! A_\mu(x)~{}^g \! A_\mu(x) \right]
\label{eq:Flogcont}
\eeq
with respect to $g(x) \in SU(3)$ acting on $A_\mu(x)$ according to  
\beq
{}^g \! A_\mu(x) =  
g(x) A_\mu(x) g^{\dagger}(x) + \frac{i}{g_0}~g(x)~\partial_\mu g^{\dagger}(x)\,.
\label{eq:gtrafocont}
\eeq
The direct transcription of the continuum extremization problem to the 
lattice leads to the minimization of a lattice gauge functional
\beq
F^{\rm (log)}_U[g] = \frac{1}{4VN_c}
\sum_{x,\mu}~\tr \left[ {}^g \! A^{\rm (log)}_{x+\frac{\hat{\mu}}{2},\mu} 
            {}^g \! A^{\rm (log)}_{x+\frac{\hat{\mu}}{2},\mu} \right]
\label{eq:Flog}
\eeq
for the logarithmic definition of the lattice gluon field [\Eq{eq:Alog}]. 
The violation of transversality then can always be checked by computing
the divergence
\bea
\Delta^{\rm (log)}(x) &=& 
\left( \sum_\mu \partial_\mu A^{\rm (log)}_\mu \right)(x) i\nonumber \\
&\equiv& 
\sum_\mu \left( A^{\rm (log)}_{x+\frac{\hat{\mu}}{2},\mu} -
                A^{\rm (log)}_{x-\frac{\hat{\mu}}{2},\mu} \right) \ne 0 \, .
\label{eq:divAlog}
\eea

Note that the logarithmic definition requires a diagonalization of the
neighboring $SU(3)$ link matrices each time the divergence
(\ref{eq:divAlog}) is evaluated.

We will see that the two lattice Landau gauge conditions 
[Eqs.\eqref{eq:divAlin} and \eqref{eq:divAlog}] --
if imposed on an arbitrary gauge field configuration -- will result in 
rather different gauge-fixed fields. However, the
gluon propagators calculated from the respective gluon fields, i.e, 
$A^{\rm (log)}_{x+\hat\mu/2,\mu}$ or $A^{\rm (lin)}_{x+\hat\mu/2,\mu}$, 
will be seen to be related to each other approximately by a finite 
multiplicative renormalization. Note also that the gluon fields transformed
into momentum space will be transversal, $q_\mu\tilde{A}_\mu(q) = 0$,  
only if the gauge-fixing procedure fits to
the respective definition of $A_{x+\hat\mu/2,\mu}$.  

\section{Landau gauge fixing: different algorithms are required}
\label{sec:algorithms}

\subsection{Linear definition}

The linear form of the gauge functional in \Eq{eq:Flin} suggests to
use a relaxation method for its minimization (Los Alamos type
gauge-fixing): at given lattice site $x$ the gauge transformation
field $g_x$ is replaced by $g^{\prime}_x$ such that the expression
\begin{equation}
  \re \tr \left\{g^{\prime}_x ~ W_x \right\}
\end{equation}
is maximized for a given $W_x$, where
\begin{displaymath}
  W_x \equiv \sum_\mu \left( U_{x,\mu}~g^{\dagger}_{x+\hat{\mu}}+ 
 U^{\dagger}_{x-\hat{\mu},\mu}~g^{\dagger}_{x-\hat{\mu}} \right)
\end{displaymath}
This is also known as a ``projection onto
$SU(3)$''  
\beq
g^{\prime}_x \equiv {\rm Proj}_{SU(3)} ~W_x^{\dagger} \,.
\eeq
For $SU(3)$ $g^{\prime}_x$ is typically found using the
Cabibbo-Marinari decomposition \cite{Cabibbo:1982zn}, such that the 
local $SU(3)$ update of $g_x$ proceeds via three successive $SU(2)$
updates. Formally, the update can be viewed as the replacement
\beq
g_x \to g^{\prime}_x \equiv r_x~g_x 
   \mbox{~~with~~} r_x=g^{\prime}_x~g^{\dagger}_x\,.
\label{eq:replacement1}
\eeq
The speed of convergence is usually improved by replacing the
relaxation steps by overrelaxation (OR) steps,
\beq
g_x \to  r_x^\omega ~ g_x\,.
\label{eq:replacement2}
\eeq
The overrelaxation parameter $\omega$ has to be optimized in the
interval $1<\omega < 2$. The required power of the 
update matrix $r_x$ is approximated via the truncated series
\beq
r^{\omega}_x = \sum_{n=0}^N~\frac{\gamma_n(\omega)}{n~!}~
                            \left( r_x - 1 \right)^n\,, 
\label{eq:OR_parameter}
\eeq
where $N=3$ or 4, and 
\beq
\gamma_n(\omega) = \frac{\Gamma(\omega+1)}{\Gamma(\omega+1-n)}\,.
\eeq

In order to check the Gribov ambiguity, the OR procedure can be
repeated for a number of initial random gauges $g^{(i)}_{\rm
initial},~i=1,\dots,N_{\rm{copy}}$ typically resulting in different
final gauge transformations $g^{(i)}_{\rm final}$ for a given
lattice gauge field $U$. The different gauge-transformed
fields are known as ``gauge copies''. They lead to
a corresponding distribution of gauge-functional values
$F^{(i)} = F^{\mathrm{(lin)}}_U[g^{(i)}]$ (the set of local minima)
instead of a unique, but usually unknown absolute minimum. 

To shift the distribution of $F^{(i)}$ to smaller values,
the initial random gauge transformation can be replaced by one
obtained using a simulated annealing algorithm (SA) for
gauge-fixing. SA is a Markov chain MC process  
simulating a Gibbs measure of the form
\beq
P_{U,T_g}[g] ~\propto~ \exp \left(-\frac{F^{\rm (lin)}_U[g]}{T_g} \right)
\eeq
for the field of gauge transformations $g$, where the gauge 
temperature $T_g$ is lowered step by step after one or a few Markov
steps according to some protocol (``SA schedule'') such  
that the process never passes through real equilibrium states. To our
knowledge, the method has been very successfully applied for the first
time fixing to maximally Abelian gauge in \cite{Bali:1996dm}. 

It turned out that with respect to computer time as well as with
respect to finding smaller $F^{(i)}$, repeating a combined SA+OR
algorithm is more efficient, than repeating the OR algorithm with 
initial random gauge transformation (see
\cite{Schemel:2006da,Bogolubsky:2007pq}). For the results presented
below we have employed
the SA+OR algorithm to gauge-fixing the fields $A^{\mathrm{(lin)}}$
[\Eq{eq:Alin}] but also to precondition the gauge-fixing of
$A^{\mathrm{(log)}}$ [\Eq{eq:Alog}] (for details see below).

For the SA algorithm we have changed the gauge temperature $T_g(i_s)$ 
for every new MC sweep ($i_s$) according to the SA schedule
\beq
T_g(i_s) = \left[ (T_{g,\min}^5 - T_{g,\max}^5 )~
\frac{i_s -1}{N_{\rm iter} -1} + T_{g,\max}^5 \right]^{\frac{1}{5}}
\eeq
proposed in \cite{Schemel:2006da} with a restriction  
to $N_{\rm iter} = 3500$ iterations. We have used the initial maximal 
gauge temperature value $T_{g,\max}=0.45$ and the final minimal value 
$T_{g,\min}=0.01$.


\subsection{Logarithmic definition}

For the logarithmic definition of the gluon field (for brevity we call the 
corresponding algorithm logarithmic gauge fixing), the starting point is the 
local divergence \Eq{eq:divAlog} evaluated at all lattice sites $x$.
The gauge transformations $g_x$ (at a given lattice site $x$) is updated
locally by $g_x \to r_xg_x$ where $r_x$ is the exponentiated
local divergence of ${}^gA_\mu$ evaluated at $x$ according to
[\Eq{eq:divAlog}], i.e.,
\beq
r_x = 
\exp\left(-i\alpha~(\sum_\mu\partial_\mu {}^g \! A^{\rm (log)}_\mu)(x)\right) \,.
\label{eq:localgradient}
\eeq
The step size $\alpha$ has to be tuned. We call this method the local
\emph{unaccelerated} steepest gradient algorithm (the original ``Cornell
type'' gauge fixing discussed in Ref.~\cite{Davies:1987vs}). 

It is well-known that this update suffers from critical slowing down
which can be ameliorated using a \emph{Fourier-accelerated} 
version~\cite{Davies:1987vs}: 
at each lattice site $g_x \to r_xg_x$ using
\beq
r_x = \exp\left(-i\alpha\hat{F}^{-1}\left[\frac{q_{\rm max}^2}{q^2} 
\hat{F}\left[(\sum_\mu\partial_\mu {}^g \! A^{\rm (log)}_\mu)(x)\right]\right] 
\right),
\label{eq:Fourier_accelerated}
\eeq
instead. Here $\hat{F}$ denotes the Fourier transformation from the space-time 
lattice to the 4-momentum lattice and $\hat{F}^{-1}$ the reverse transformation,
respectively (for $q^2$ see (\ref{eq:mom1}) and (\ref{eq:mom2})).
One recognizes 
a
sort of smearing of the divergence with the inverse Laplacian. 
We notice that Furui and Nakajima~\cite{Nakajima:2000yf} have also used a 
non-local algorithm in the form of a Newton-Raphson-type construction of 
$r_x$ using the non-vanishing divergence as a source and the inverse Hessian
(Faddeev-Popov operator, expanded in $A_\mu$ to some finite order) in place 
of the inverse Laplacian.

Fast Fourier transformations are less efficient if the lattice is fully 
parallelized. This could become a problem for large lattices. We
circumvented this problem by using the \emph{multigrid-accelerated} steepest
gradient algorithm \cite{Goodman:1986pv}. In fact, the last
expression for $r_x$ can be written as
\beq
r_x = \exp\left(-i\alpha\,q_{\rm max}^2~\Delta^{-1} 
(\sum_\mu \partial_\mu {}^g \! A^{\rm (log)}_\mu)(x)\right).
\label{eq:Multigrid_accelerated}
\eeq
The inversion of the Laplacian on an arbitrary source
is a standard problem for a multigrid algorithm. Details on our
implementation of the multigrid-accelerated gradient algorithm are
given in the Appendix.

We faced the problem that the logarithmic gauge fixing fails to finish 
successfully for 10\% to 50\% of all attempts, when starting
from a random gauge transformation. The exact failure rate depends on
the gauge-fixing method, and grows with the lattice size. A similar
problem only rarely occurs in case of the linear gauge fixing using
the standard OR algorithm. We found, however, that the logarithmic
algorithm works successfully in 99\%  of the cases if the divergence
[\Eq{eq:divAlog}] is already sufficiently small at the start. This can be 
achieved by preconditioning the logarithmic by a linear gauge fixing.
The remaining 1\% of cases is dealt with by just repeating the gauge
fixing.

\begin{figure*}[tb]
        \centering
        \mbox{
        \includegraphics[angle=0]{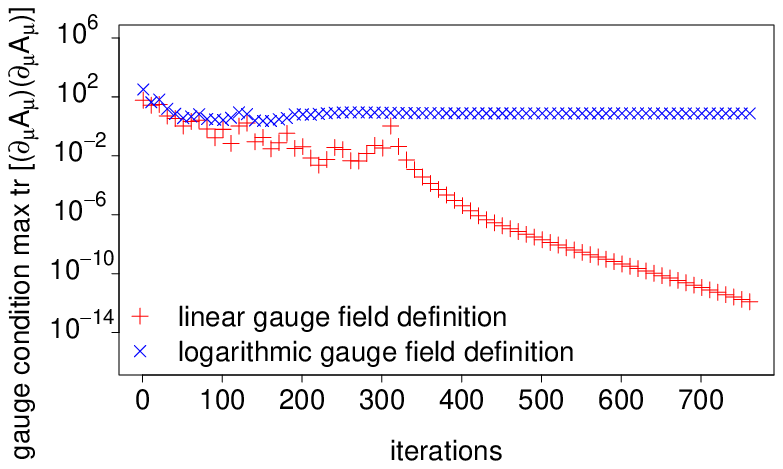} 
        \qquad 
        \includegraphics[angle=0]{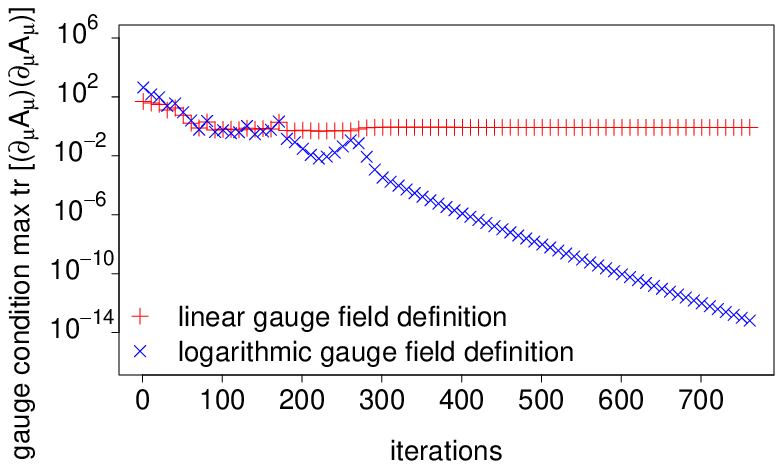} 
         }
\caption{
   History of maximal values of the squared divergence 
   of the two gluon field definitions during gauge fixing by
   overrelaxation (left) and the Fourier-accelerated gradient
   method (right). Example configuration on a $16^4$ lattice at 
   $\beta = 6.0$.}
\label{fig:figure1}
\medskip
        \mbox{
        \includegraphics[angle=0]{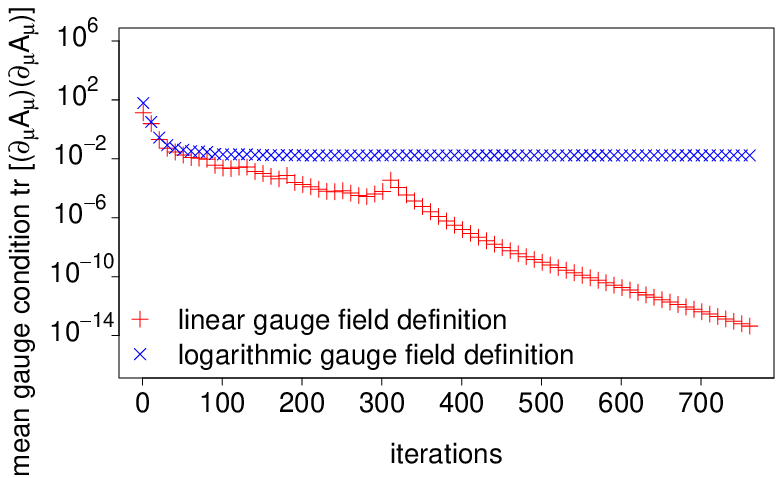} 
        \qquad 
        \includegraphics[angle=0]{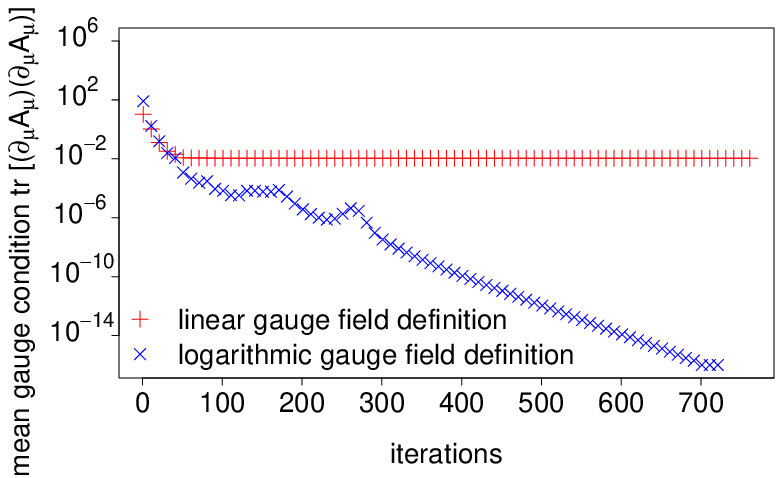} 
        }
\caption{
   Same as in \Fig{fig:figure1} but for the space-time average of the
   squared divergence of the respective $A_\mu$.}
\label{fig:figure2}
\end{figure*}


\subsection{Comparing the two gauge fixing prescriptions}

For both types of gauge fixing a stopping criterion is needed. This
criterion is fulfilled as soon as the respective divergence,
$\left( \partial_\mu {}^g \! A^{\rm (lin)}_\mu \right)(x)$ or
$\left( \partial_\mu {}^g \! A^{\rm (log)}_\mu \right)(x)$ is sufficiently small.
We have applied the criterion
\beq
\max_x~\tr~\left[ \left( \partial_\mu {}^g \! A_\mu \right)^{\dagger}(x)
           \left( \partial_\mu {}^g \! A_\mu \right)(x) \right] < 10^{-14}\,.
\eeq

To demonstrate the relation between the two different gauge-fixing
algorithms we compare now the history of the overrelaxation and of the
Fourier-accelerated gradient algorithm. Both are applied to the same
configuration on a $16^4$ lattice (generated using the Wilson
plaquette action at $\beta = 6.0$).

We used $\omega = 1.68$ and $N = 3$ [see \Eq{eq:OR_parameter}] for the 
linear gauge-fixing method (in this case the OR algorithm without the
SA-preconditioning step) and $\alpha = 0.07$ 
[\Eq{eq:Fourier_accelerated}] for the logarithmic gauge fixing 
(i.e., the Fourier-accelerated gradient algorithm).

In \Fig{fig:figure1} we compare how the local maxima of the squared
divergence of $A^{\rm (lin)}$ and $A^{\rm (log)}$ behave while either the
overrelaxation (left panel) or Fourier-accelerated gradient algorithm
(right panel) proceeds. When the OR algorithm fixes the Landau-gauge
the divergence of $A^{\rm (lin)}$ is successively reduced. This
cannot be simultaneously achieved by $A^{\rm (log)}$.
On the other hand, when the gradient algorithm fixes the Landau-gauge the
divergence of $A^{\rm (log)}$ becomes smaller with every iteration,
while the divergence of $A^{\rm (lin)}$ stays almost unchanged. We notice
that the local maximum of the squared divergence of the ``wrong gauge
field'' never becomes less than ${\cal O}(1)$. This is the result of
only few isolated defects, as it can be seen by a comparison with the
corresponding space-time averaged quantities (\Fig{fig:figure2}). The
space-time average defined through $||\Delta||^2$ [Eq.~\eqref{eq:deltanorm}] 
reaches a precision of $10^{-4}$, typically after
${\cal O}(100)$ or ${\cal O}(200)$ iterations. To reach the same
precision for all lattice sites more iterations are needed. We notice
that the space-time average of the squared divergence of the ``wrong
gauge field'' never becomes less than ${\cal O}(10^{-2})$.
Comparing the linear and the quadratic definition of the gluon field
the authors of Refs.~\cite{Giusti:1998ur,Petrarca:1998je} have observed 
a similar difference of  $||\Delta^{\rm (quad)}||^2$ and 
$||\Delta^{\rm (lin)}||^2$.

Local operators constructed in terms of the gauge-fixed gluon fields 
can only be sufficiently precise if the respective gauge-fixing method
has been used for each $A_\mu$. However, as a preconditioner we can
(and we do) use the linear gauge-fixing (via the SA+OR algorithm)
before applying logarithmic gauge fixing. As mentioned above, this
helps much to reduce the number of unsuccessful gauge-fixing attempt
for the latter. For the linear definition the SA+OR algorithm always 
converges.

\begin{figure*}
        \centering
        \mbox{
        \includegraphics[angle=0]{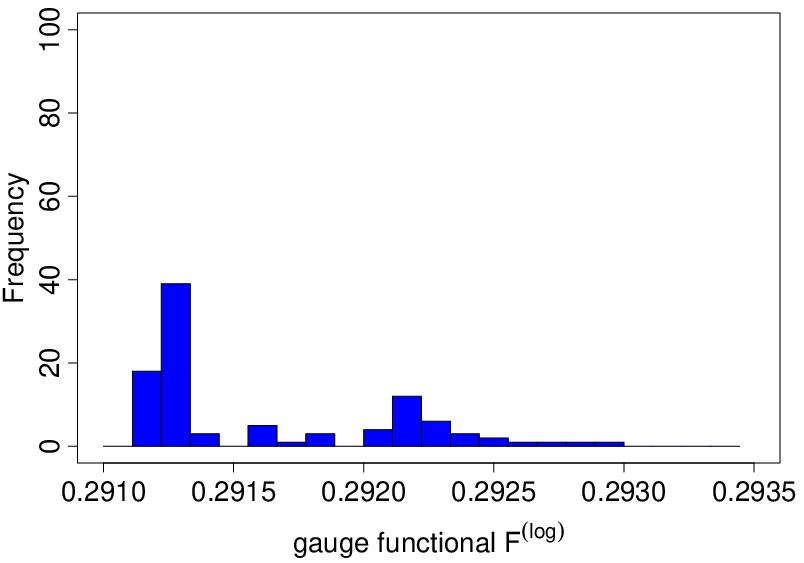}
        \qquad 
        \includegraphics[angle=0]{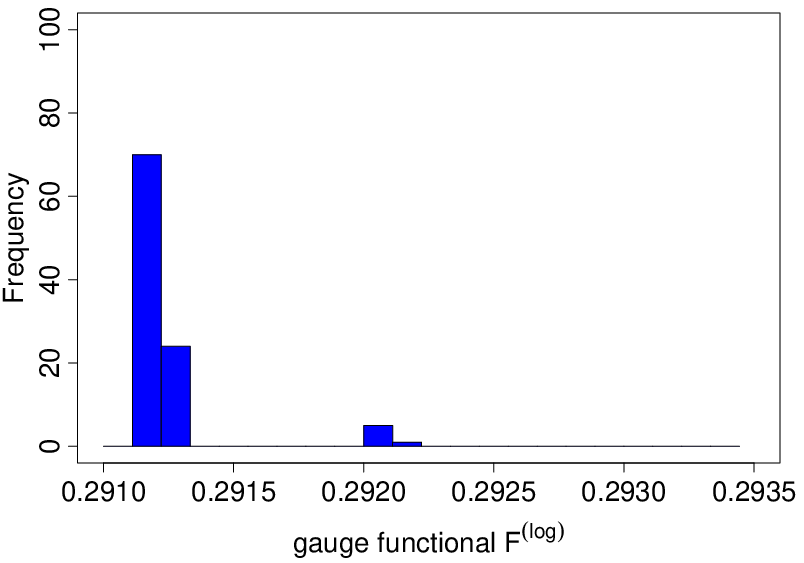}
        }
\caption{
   The effect of replacing OR (left) by SA+OR (right) as preconditioner.
   The figure shows a histogram of final $F^{\rm (log)}$ values for
   100 gauge-fixed copies ($16^4$ lattice, $\beta = 6.0$). The
   multigrid-accelerated gradient method was used for the gauge-fixing.} 
\label{fig:figure3}
\medskip
        \centering
        \mbox{
        \includegraphics[angle=0]{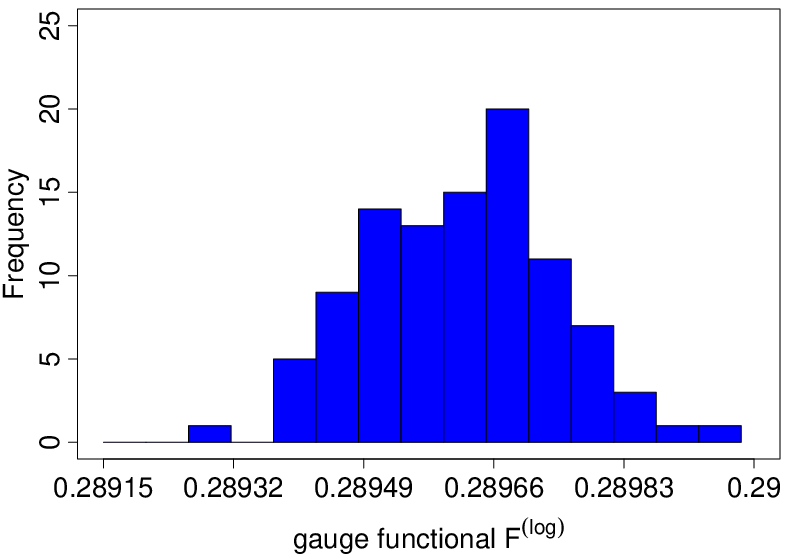}
        \qquad 
        \includegraphics[angle=0]{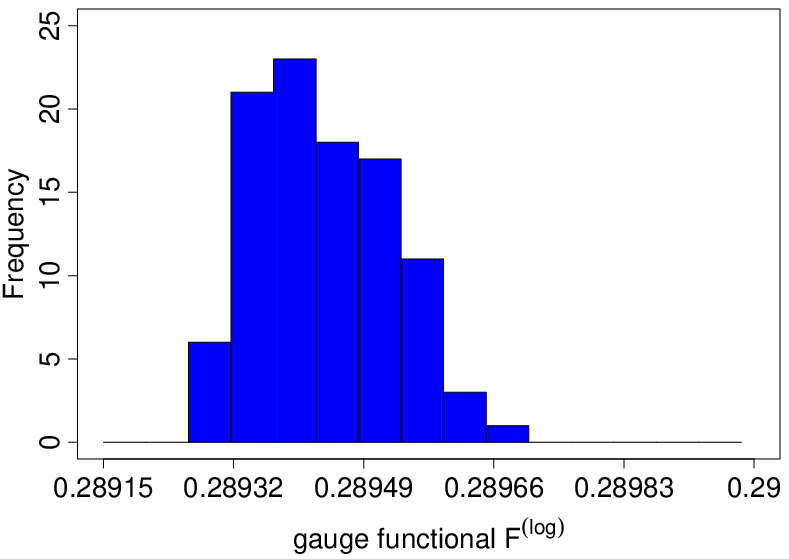} }
\caption{Same as \Fig{fig:figure3}, but for a $32^4$ lattice.} 
\label{fig:figure4}
\end{figure*}

Preconditioning is not without effect on the final gauge fixing,
however. The quality of gauge fixing achieved by the (linear)  
preconditioner influences the quality of the final (logarithmic) 
gauge fixing. This is illustrated in Figs.~\ref{fig:figure3} and
\ref{fig:figure4} at $\beta=6.0$ for a $16^4$ and $32^4$ lattice, respectively.
%
These figures show the effect of replacing pure OR by SA+OR 
(with respect to $A^{\rm (lin)}$) as preconditioner for the logarithmic
gauge fixing (performed with the multigrid-accelerated gradient method) 
on the distribution of the final gauge functional values
$F^{\rm (log)}_U[g^{(i)}]$. This comparison is again for a
$16^4$ lattice at $\beta=6.0$.

As well-known for the linear gauge fixing, we typically find
smaller gauge-functional values also for the logarithmic case, when  
the SA+OR algorithm is used instead of the OR algorithm for this
preconditioning step. This becomes even more pronounced increasing the
lattice size where the number of local minima naturally increases.


\subsection{Performance of different realizations of the logarithmic 
gauge fixing}

We compare now the logarithmic gauge fixing (without SA+OR 
preconditioning) in three versions: the unaccelerated gradient
algorithm, the Fourier-accelerated gradient algorithm and the
multigrid-accelerated gradient algorithm. The advantage of the latter
is that it is easy parallelizable. Our code employes a further developed
version of the algorithm used by Cucchieri and Mendes for  
Landau gauge in $SU(2)$ gauge theory \cite{Cucchieri:1998ew}.
In the Appendix our implementation of the multigrid-accelerated
gradient algorithm is described in more detail.  

\begin{figure*}
        \centering
        \mbox{
        \includegraphics[angle=0]{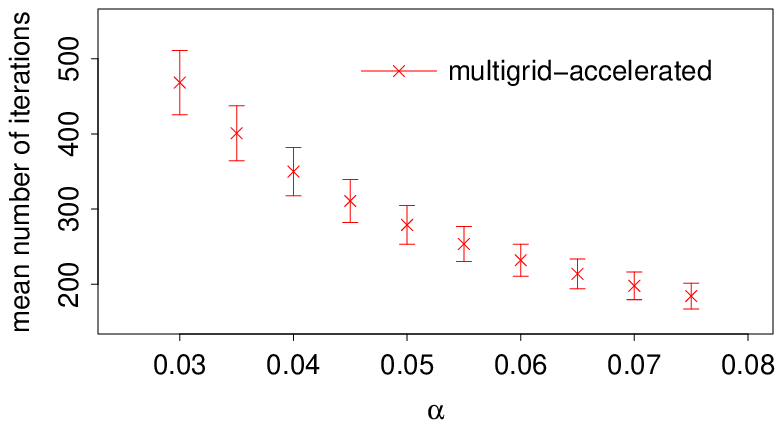}   
        \qquad 
        \includegraphics[angle=0]{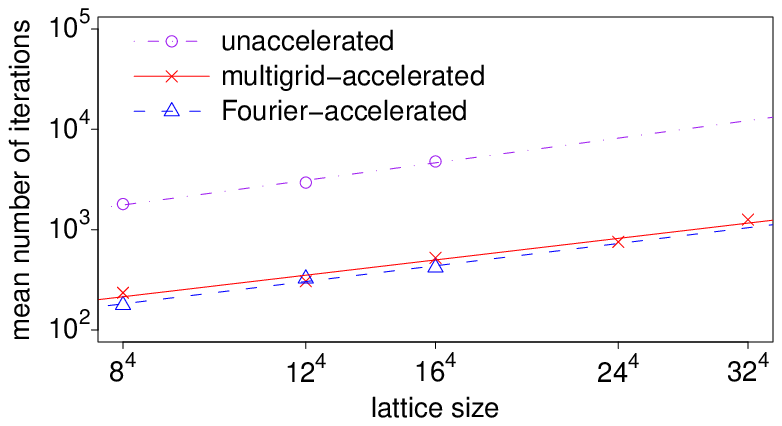}  }
        \caption{
        The number of iterations needed to reach the stopping 
        criterion. Left: as a function of the step size parameter $\alpha$ of  
        the multigrid-accelerated algorithm (lattice size $16^4$, $\beta = 6.0$). 
        Right: as a function of the lattice size for the three investigated 
        logarithmic gauge fixing algorithms ($\beta = 6.0$).
        }
\label{fig:figure5}
\end{figure*}
In \Fig{fig:figure5} (left) we show the average number of iterations of 
the multigrid-accelerated algorithm as a function of the step size parameter 
$\alpha$ (again for a $16^4$ lattices with $\beta=6.0$).
With increasing $\alpha$ the mean number of iterations decreases monotonously 
until it reaches an optimal value $\alpha = \alpha_{\mathrm{opt}}$.
Further increasing $\alpha$ beyond that value leads to
instabilities and is therefore avoided. In \Tab{tab:table1} we summarize
our values on $\alpha_{\mathrm{opt}}$, for the three algorithms and
different lattice sizes and $\beta$.
Due to limited computing resources, for the bigger lattices we could 
afford to fix the gauge only by means of the 
multigrid-accelerated gradient algorithm (in its parallelized  
version).\footnote{The lattice ensembles at $\beta=9.0$ have only been used 
for comparison with NSPT. Larger lattice sizes could be simulated by NSPT only 
quite recently. They were not available at the time when the MC studies 
described here were finished.}
%
%
\begin{table}[b]
\centering
  \caption{Optimal values $\alpha_\text{opt}$ for the three different
    logarithmic gauge-fixing algorithms for $\beta = 6.0$ and 9.0, and
    different lattice sizes. Due to the high demand of computation time,
    $\alpha_\text{opt}$ has not been determined for the unaccelerated and
    the Fourier-accelerated algorithm on lattices larger than $16^4$.}  
  \label{tab:table1}
\begin{tabular}{r@{\quad}c@{\quad}c@{\quad}c@{\quad}ccc} 
\hline\hline
algorithm             & $\beta$ & $8^4$   & $12^4$  & $16^4$  & $24^4$ & $32^4$ \\
\hline
unaccelerated         & $6.0$   & $0.130$ & $0.130$ & $0.110$ &    -   &    -   \\
Fourier-accelerated   & $6.0$   & $0.070$ & $0.070$ & $0.070$ &    -   &    -   \\
multigrid-accelerated & $6.0$   & $0.070$ & $0.070$ & $0.075$ & $0.70$ & $0.70$ \\*[1ex]
unaccelerated         & $9.0$   & $0.125$ & $0.125$ & $0.125$ &    -   &    -   \\
Fourier-accelerated   & $9.0$   & $0.065$ & $0.065$ & $0.065$ &    -   &    -   \\
multigrid-accelerated & $9.0$   & $0.065$ & $0.065$ & $0.075$ &    -   &    -   \\
\hline\hline
\end{tabular}
\end{table}
%
\Fig{fig:figure5} (right) presents the scaling of the average iteration 
number with the lattice size for the three logarithmic gauge-fixing algorithms 
with their respective $\alpha_\text{opt}$ (again all for $\beta=6.0$). 
We find that these numbers to scale like
\beq
N_{\rm iter} = C~V^z\,.
\label{eq:volume}
\eeq
Values for $C$ and $z$ are summarized in \Tab{tab:table2}.

Apparently, the two accelerated algorithms perform much better
compared to the unaccelerated one, which is mainly due to a much
smaller $C$ [compare $\mathcal{O}(10)$ versus $\mathcal{O}(100)$]. The 
values for $z$ are almost the  same for all the three algorithms. We
favor the multigrid gauge-fixing algorithm because it is easy to
parallelize.

%
\begin{table}[bt]
\centering
 \caption{The exponent and coefficient, $z$ and $C$, of \Eq{eq:volume}
for the three different logarithmic gauge-fixing algorithms.} 
\label{tab:table2}
\begin{tabular}{r@{\qquad\qquad}r@{\qquad\qquad}c} 
\hline\hline
algorithm             & C     & z         \\
\hline
unaccelerated         & $110$ & $0.33(4)$ \\
Fourier-accelerated   & $13$  & $0.3(4)$  \\
multigrid-accelerated & $11$  & $0.3(4)$  \\
\hline \hline
\end{tabular}
\end{table}


\section{Gluon and ghost propagators for $A_\mu^{\rm (lin)}$ 
and $A_\mu^{\rm (log)}$}
\label{sec:compareobservabledefinitions}

\subsection{Gluon propagator}

We are interested in the gluon and ghost propagators for the linear
and logarithmic definition. With the lattice gluon field $A^{\rm (lin)}$ and
$A^{\rm (log)}$, respectively, the bare gluon propagator is defined as
\beq 
D^{ab}_{\mu,\nu}(x,y) = 
\left\langle A^a_{x+\frac{\hat{\mu}}{2},\mu} A^b_{y+\frac{\hat{\nu}}{2},\nu}\right\rangle_U,
\eeq 
where ${\langle ... \rangle}_U$ denotes the ensemble average over gauge-fixed 
configurations. As in most of the applications we evaluate it within
the lattice Fourier representation
\beq 
\tilde{D}^{ab}_{\mu\nu}(q(k)) = 
\frac{1}{V}~\left\langle \tilde{A}^a_{\mu}(k)~\tilde{A}^b_{\nu}(-k) \right\rangle_U
\eeq
with 
\beq
\tilde{A}^a_{\mu}(k)=
\sum_x~A^a_{x+\frac{\hat{\mu}}{2},\mu}{\rm e}^{ik\cdot(x+\frac{\hat{\mu}}{2})}\,, 
\eeq
where the abbreviation $k \cdot x = \sum_\mu \frac{2\pi k_\mu x_\mu}{L_\mu}$
has been used. $L_\mu$ denotes the lattice size in
$\mu$-direction. The integers $k_{\mu} \in (-L_\mu/2, +L_\mu/2]$ count
the momentum modes within the Brillouin zone. The lattice momenta can
be written in two forms, 
\beq 
\overline{q}_\mu(k) = \frac{2 \pi k_\mu}{a L_\mu}
\label{eq:mom1}
\eeq
and
\beq 
q_\mu(k) = \frac{2}{a} \sin \frac{\pi k_\mu}{L_\mu} = 
\frac{2}{a} \sin \frac{a \overline{q}_\mu}{2}\,.
\label{eq:mom2}
\eeq
The latter is the one that appears in the lattice tree-level
expression for the gluon propagator, and therefore taken as the
corresponding physical momentum.

For the lattice spacing dependence $a(\beta)$ we adopt
\cite{Necco:2001xg} and use $r_0=0.5$\,fm to assign physical units to $a$.
\Tab{tab:table3} lists the lattice spacing values used in this
study.

%
\begin{table}[t]
\centering
 \caption{Values for the lattice spacing $a(\beta)$ and its inverse
   as used in this study. We used the formula given in \cite{Necco:2001xg}
   with $r_0 = 0.5 \mathrm{~fm}$.} 
\label{tab:table3}
\begin{tabular}{c@{\qquad\qquad}c@{\qquad\qquad}c} 
\hline \hline
$\beta$         & $a(\beta)$ in fm & $a^{-1}(\beta)$ in GeV         \\
\hline
$5.8$           & $0.1364$             & $1.4464$ \\
$6.0$           & $0.0932$             & $2.1184$ \\
$6.2$           & $0.0677$             & $2.9137$ \\
$6.4$           & $0.0513$             & $3.8445$ \\
\hline
\end{tabular}
\end{table}

Supposed all non-diagonal components in the color ($a,b$) and the
Euclidean indices ($\mu,\nu$) vanish, one can average
$\tilde{D}^{ab}_{\mu\nu}$ over the diagonal elements 
\beq 
D(q^2) \equiv 
\frac{1}{8} \sum_a \frac{1}{3} \sum_\mu \tilde{D}^{aa}_{\mu\mu}(q(k))\,.
\eeq
The factor $1/3$ is due to the transversality of the gluon field with respect 
to the vector $q_\mu(k)$ that leaves only three independent modes. 
The gluon dressing function is
\beq 
Z_{Gl}(q^2)= q^2 D(q^2)\,.    
\eeq


\subsection{Ghost propagator}

For the ghost propagator the situation is somewhat different.
It is the two-point function of the ghost fields $c^a$ 
and $\overline{c}^b$, and these fields are only implicitly defined
through the Faddeev-Popov operator $M^{ab}_{xy}$. 
Therefore, the bare ghost propagator is defined as the ensemble average 
of the inverse Faddeev-Popov operator, i.e.,
\beq 
G^{ab}(x,y) = \langle c^a(x)~\overline{c}^b(y) \rangle = 
\left\langle \left( M^{-1} \right)^{ab}_{xy} \right\rangle_U\,.
\eeq

As mentioned the form of the Faddeev-Popov operator depends on the
definition adopted for the gluon field. Generally, on the  
lattice one can decompose the Faddeev-Popov operator as follows: 
\beq 
M^{ab}_{xy} = A^{ab}_x~\delta_{xy} 
- \sum_\mu \left(B^{ab}_{x,\mu}~\delta_{x+\hat{\mu},y} 
+ C^{ab}_{x,\mu}~\delta_{x-\hat{\mu},y} \right)\,.
\label{eq:FP}
\eeq
For the linearly defined gluon fields one determines the Faddeev-Popov operator 
as the Hessian of the gauge functional $F^{\rm (lin)}$ with respect to 
infinitesimal gauge transformations. This leads to the following form of the 
adjoint representation matrices entering (\ref{eq:FP}) 
\begin{eqnarray} 
A^{ab}_x       & = &   
\re \tr \left[ \{T^a,T^b\} 
\sum_\mu \left( U_{x,\mu} + U_{x-\hat{\mu},\mu} \right) \right] \,, \nonumber \\
B^{ab}_{x,\mu} & = & 2~\re \tr \left[ T^b~T^a~U_{x,\mu} \right] \,, \\
C^{ab}_{x,\mu} & = & 2~\re \tr \left[ T^a~T^b~U_{x-\hat{\mu},\mu} 
\right] \,, 
\nonumber\label{eq:EQ32} 
\end{eqnarray}
where $T^a$ ($a=1,\ldots,8$) denote the generators of $SU(3)$.

For the logarithmic definition the following form of the adjoint representation 
matrices is used,
\begin{eqnarray} 
A^{ab}_x & = & \sum_\mu \left[\Omega^{ab}_{x-\hat{\mu},\mu}+\Omega^{ab}_{x,\mu}    
              - A^c_{x+\frac{\hat{\mu}}{2},\mu}~f^{abc} \right] \,, \nonumber \\
B^{ab}_{x,\mu} & = & \Omega^{ab}_{x,\mu} \,, \\
C^{ab}_{x,\mu} & = & \Omega^{ab}_{x-\hat{\mu},\mu}  
                        -  A^c_{x-\frac{\hat{\mu}}{2},\mu}~f^{abc} \,, \nonumber 
\label{eq:EQ33}
\end{eqnarray}
where (written up to fourth order in the gluon field) 
\begin{eqnarray} 
\Omega^{ab}_{x,\mu} & = & 
\Biggl( \frac{i~A_{x+\frac{\hat{\mu}}{2},\mu}} {1 - U^{\dagger}_{x,\mu}} 
\Biggr)^{ab} 
\nonumber \\
 & \approx &  \delta^{ab} 
+ \frac{i}{2}~A^{ab}_{x+\frac{\hat{\mu}}{2},\mu} -
\\ \nonumber
&~~&
- \frac{1}{12}~\left( A^2_{x+\frac{\hat{\mu}}{2},\mu} \right)^{ab} -  
\frac{1}{720}~\left( A^4_{x+\frac{\hat{\mu}}{2},\mu} \right)^{ab}
\label{eq:EQ34}
\end{eqnarray}
with the gluon field $A^{ab}$, its square $\left( A^2 \right)^{ab}$ etc. 
taken in the adjoint representation.

The Faddeev-Popov operator is inverted with a Laplacian-preconditioned 
conjugate gradient algorithm and a color-diagonal plane wave source
as explained in~\cite{Sternbeck:2005tk}.

For both the linear [\Eq{eq:EQ32}] and the logarithmic definition
[\Eq{eq:EQ33}] the ghost propagator in the momentum space is given by
[$q=q(k)$] 
\beq
G^{ab}(q^2) = \frac{1}{V} \sum_{x,y} \left\langle {\rm e}^{-2\pi i\,k
      \cdot (x-y)} [M^{-1}]^{ab}_{xy} \right\rangle_U = \delta^{ab} G(q^2)\,.
\label{eq:ghostprop}
\eeq
The corresponding ghost dressing function is
\beq 
Z_{Gh}(q^2)= q^2 G(q^2)\,.    
\eeq


\begin{figure*}
        \centering
        \mbox{
        \includegraphics[angle=0]{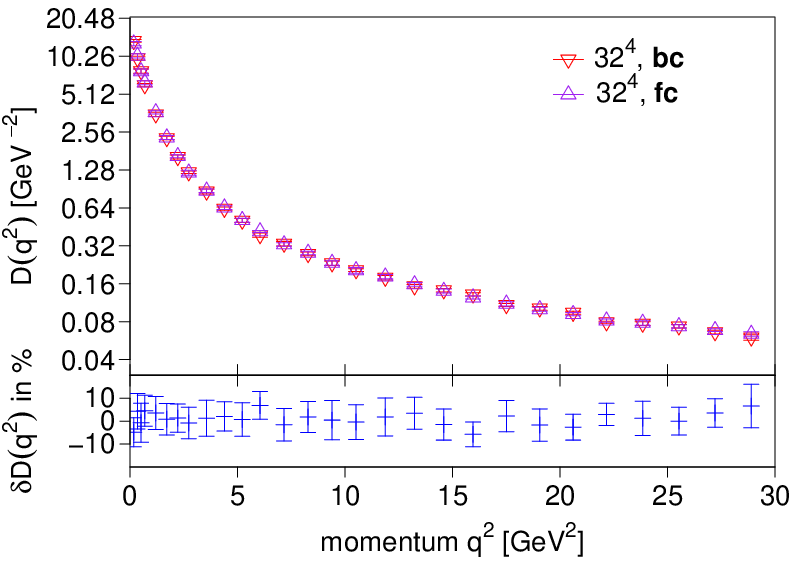}
        \qquad 
        \includegraphics[angle=0]{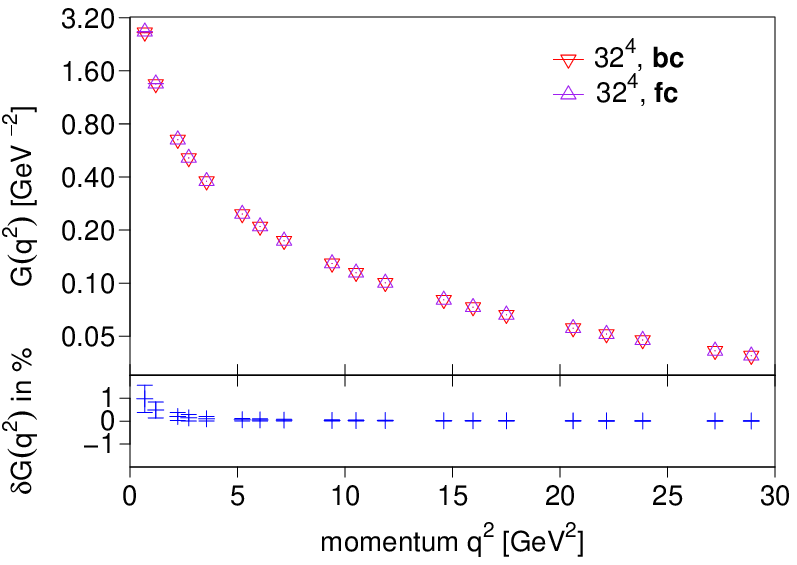} }
        \caption{Left: the bare gluon propagator $D(q^2)$ from the 
	logarithmic gluon field definition and the relative difference 
	$\delta D(q^2)$ in percent between $\fc$- and $\bc$-configurations generated 
	at $\beta = 6.0$ for lattice size $32^4$. Right: the corresponding
	bare ghost propagator $G(q^2)$ with its relative difference
	$\delta G(q^2)$.}
\label{fig:figure6}
\end{figure*}

\subsection{Momentum cuts}

On the lattice, the $O(4)$-symmetry of the continuum Euclidean space-time 
is broken to the discrete $H(4)$-symmetry. In order to minimize the
resulting artifacts, due to the finite lattice volume and spacing we
apply two momentum   
cuts: a \emph{cone} and a 
\emph{cylinder cut}~\cite{Leinweber:1998uu}.~\footnote{For 
an alternative approach see, e.g., Ref.~\cite{deSoto:2007ht}.}
The cone cut removes all (low) lattice momenta 
$k=(k_1,k_2,k_3,k_4)$ with at least one vanishing $k_i$. 
It is applied to minimize finite-volume effects associated with these
lattice momenta. The cylinder cut removes all momenta which are not
close to a multiple of one of the space-time diagonal unit vectors  
$n=(1/2)(\pm 1,\pm 1,\pm 1,\pm 1)$. For a symmetric lattice, this
criterion can be formulated as
\beq
\sum^4_{i=1} k^2_i - \left[ \sum^4_{i=1} k_i n_i \right]^2 \le 1 \; .
\label{eq:cylinder_cut}
\eeq
These two cuts removed most of the lattice artifacts from the
data. The remaining artifacts can be assessed using different
lattice spacings $a$ and lattice volumes $V$.


\subsection{Gribov ambiguity for $A_\mu^{\rm (log)}$-propagators}

In order to carry out a first check of the Gribov ambiguity of the gluon 
and ghost propagators for the logarithmic definition $A_\mu^{\rm (log)}$ 
we used $100$ ($30$) configurations for a $16^4$ ($32^4$) lattice
(again all for $\beta = 6.0$). For the larger lattice 
we expect a stronger influence of the actual selection among the gauge-fixed 
copies. 

Employing the SA+OR preconditioned, multigrid-accelerated gradient 
algorithm we fix the gauge for each configuration $10$ times starting from 
different random initializations of the gauge transformation field $g_x$. 
As in \cite{Bakeev:2003rr} for each configuration we have calculated the 
gluon- and ghost propagator for the first (random) gauge-fixing attempt 
(first copy, $\fc$) and for the gauge field with the lowest gauge functional 
value achieved (best copy, $\bc$). \Fig{fig:figure6} shows the gluon 
propagator $D(q^2)$ and the ghost propagator $G(q^2)$, 
respectively, averaged over the first and best copies, respectively. 
In the lower panels of the figures we present also the corresponding relative 
differences $\delta D$~and~$\delta G \equiv$ ``($(\fc - \bc)/\bc$)'' of 
$\fc$- and $\bc$-results given in percent.

The gluon propagator does not show any effect of the Gribov ambiguity beyond 
statistical noise (``Gribov noise'') over the whole momentum range, while 
the ghost propagator seems to exhibit a slight systematic shift within 
the low momentum region. This shift is approximately 
$1\%$ for the $32^4$ lattice while for the $16^4$ lattice it turns out
to be negligible. The small effect of the Gribov ambiguity is certainly
a consequence of the preconditioning step for which the SA+OR
algorithm has been employed. Whether it becomes more enhanced, 
when taking global $Z(3)$-flip transformations into account
\cite{Bogolubsky:2007bw, 
Bornyakov:2008yx,Bornyakov:2009ug}, remains to be seen.

In the following we will always rely on $\bc$-results.

\section{Monte Carlo results for the propagators}
\label{sec:MCresults}

\subsection{Comparing propagator results for \hbox{$A_\mu^{\rm (lin)}$ and $A_\mu^{\rm (log)}$}}
\label{sec:compareresultslogversuslin}

To get better insight into discretization effects we compare now the
gluon and ghost dressing functions for the logarithmic and linear
definition. Unfortunately though, we have to restrict the discussion to  
relatively small lattice sizes.

%
\begin{table}[bt]
 \caption{Statistics of Monte Carlo ensembles. $N_{\mathrm{conf}}$
   gives the number of analyzed configurations and $N_{\mathrm{copy}}$
   of inspected gauge-copies for each. Values are the same for all $\beta$.}
\label{tab:table4}
\begin{tabular}{c@{\qquad\qquad\qquad}c@{\qquad\qquad\qquad}c} 
\hline\hline
lattice & $N_{\mathrm{conf}}$ & $N_{\mathrm{copy}}$ \\
\hline
$12^4$ & 200  & 10\\
$16^4$ & 100  & 10\\
$24^4$ & 50   & 10 \\
$32^4$ & 30   & 10 \\
\hline\hline
\end{tabular}
\end{table}
The data presented are based on ensembles of gauge field configurations
with statistics as given in \Tab{tab:table4}.

\begin{figure*}[tb]
   \centering
   \mbox{
   \includegraphics[angle=0]{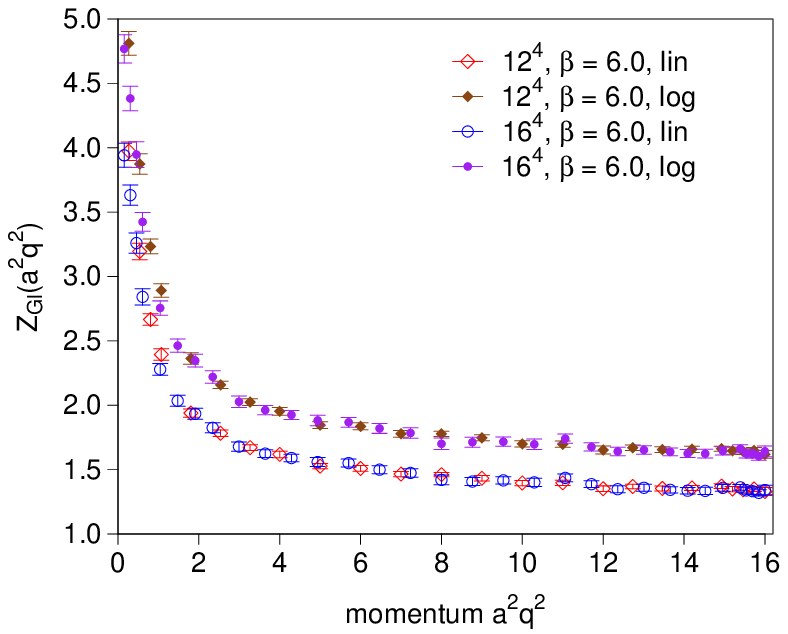}
   \qquad 
   \includegraphics[angle=0]{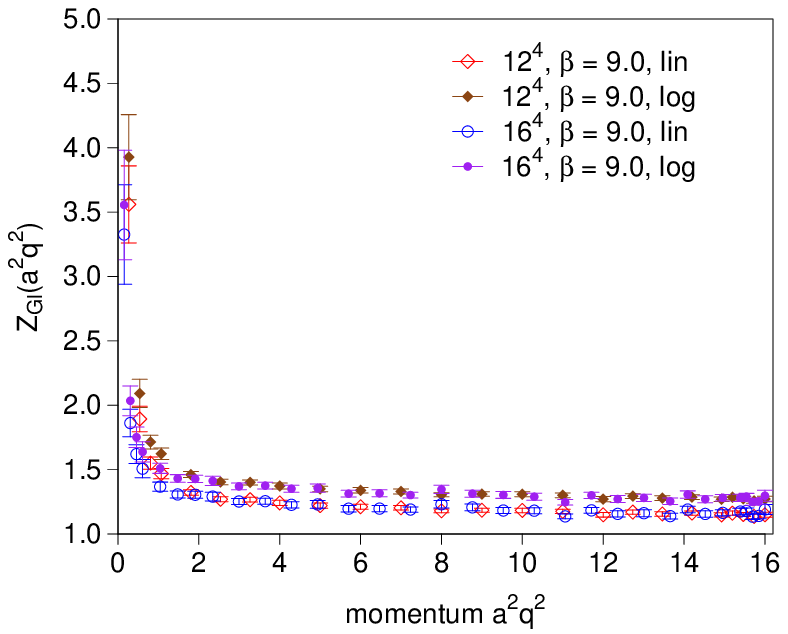}  }
\caption{The bare gluon dressing function for 
   $\beta=6.0$ (left) and $\beta=9.0$ (right) versus the lattice
   momentum squared. The data is for two lattice sizes. Filled symbols
   are for the logarithmic definition and open symbols for the linear.} 
   \label{fig:figure7}
\end{figure*}

We consider first the gluon dressing function calculated for the
linear and logarithmic definition on a $12^4$ and $16^4$ lattice
at $\beta = 6.0 $ and $\beta=9.0$. Note that the latter was chosen
only in order to compare with available NSPT results (see  
\Sec{sec:compareresultslogversusNSPT}). \Fig{fig:figure7} shows the
data for the bare dressing function versus the lattice
momentum squared. We clearly see the expected momentum-independent
offset between the results for the logarithmic and the linear
definition.

\begin{figure*}
   \centering
   \mbox{
   \includegraphics[angle=0]{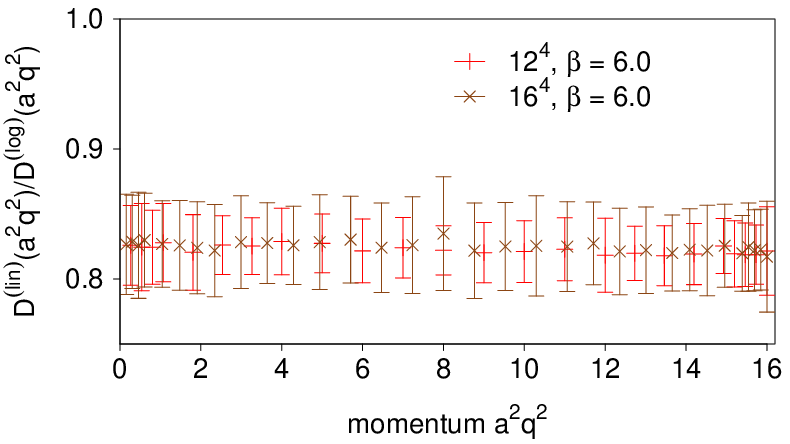}
   \qquad 
   \includegraphics[angle=0]{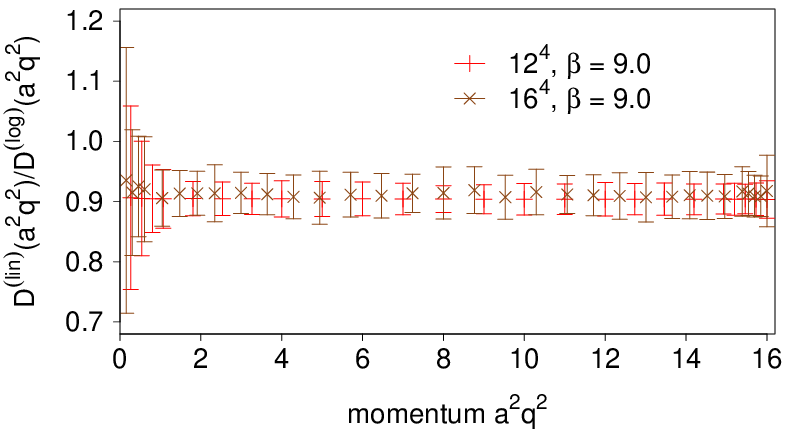}  }
\caption{The ratio $C_{\rm gluon}$ acc.\ to \eqref{eq:gluon_renorm_const} 
   relating the gluon propagator for the two definitions of the gluon
   field. Data is for $\beta = 6.0$ (left) and 
   $\beta = 9.0$ (right) on a $12^4$ and $16^4$ lattice.}
   \label{fig:figure8}
\end{figure*}

This is even better seen in \Fig{fig:figure8} where the ratio
\beq
  C_{\rm gluon}(q^2)=\frac{D^{\rm (lin)}(q^2)}{D^{\rm (log)}(q^2)} 
 \label{eq:gluon_renorm_const}
\eeq
is shown versus momentum. For both $\beta$-values, we observe the
ratio to be constant within statistical errors over the whole momentum
region, and to depend on $\beta$. Thus the bare gluon propagators
differ for the two definitions but are related to each other 
by a finite $\beta$-dependent multiplicative renormalization constant. 
As a consequence both definitions will lead to the same propagator
when renormalized in a MOM scheme, the latter being defined by the
condition that the propagator equals its tree-level expression
at some subtraction momentum $q^2=\mu^2$. 
Of course, this multiplicative renormalizability has been numerically 
demonstrated here only for finite volume and corresponding 
restricted momentum range under consideration. 
%

\begin{figure*}
   \centering
   \mbox{
   \includegraphics[angle=0]{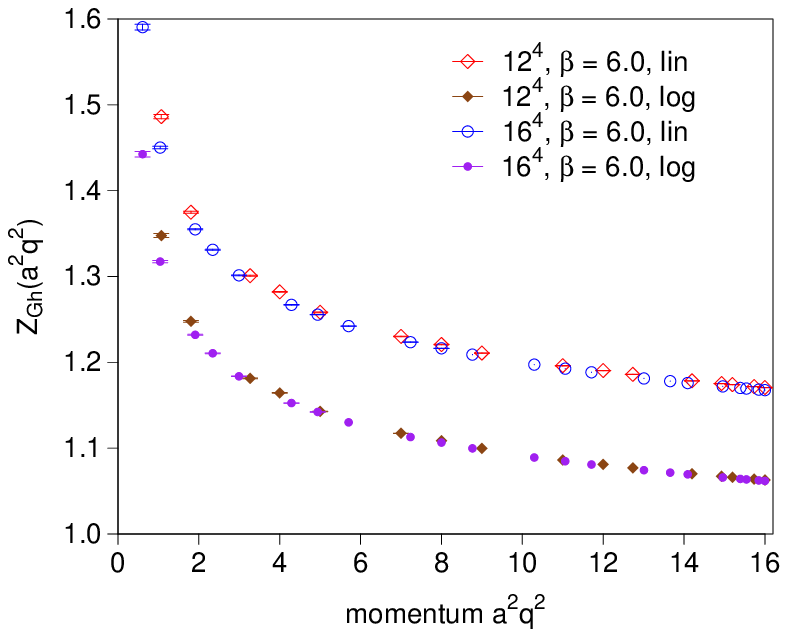}
    \qquad 
   \includegraphics[angle=0]{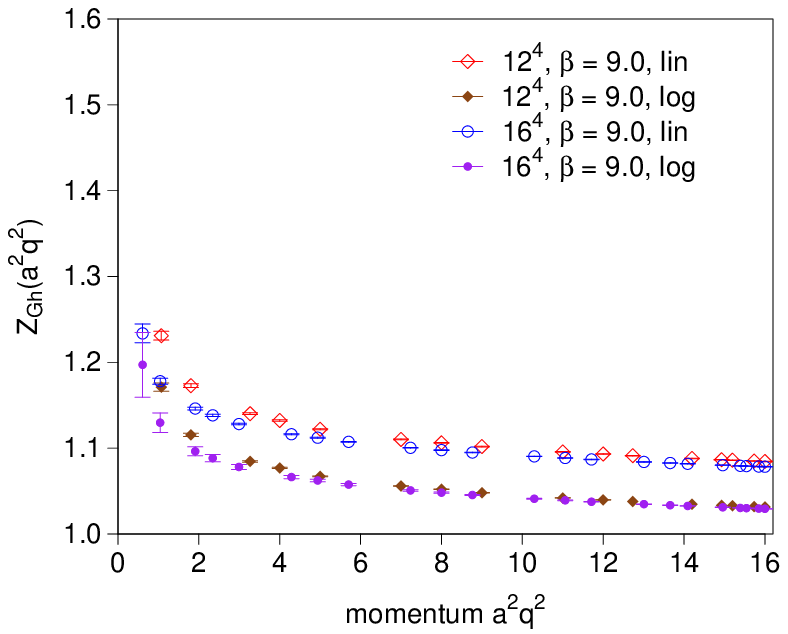}  }
\caption{The bare ghost dressing functions versus the lattice
  momentum squared. Data is for $\beta=6.0$ (left) and
  $\beta=9.0$ (right) and two lattice sizes. Filled (open) symbols are
  for the logarithmic (linear) definition.} 
\label{fig:figure9}
\end{figure*}

Also for ghost propagator we clearly see the offset between the
dressing functions for the logarithmic and linear definitions (see
\Fig{fig:figure9} for the data at $\beta = 6.0$ and $9.0$).

\begin{figure*}[tb]
   \centering
   \mbox{
   \includegraphics[angle=0]{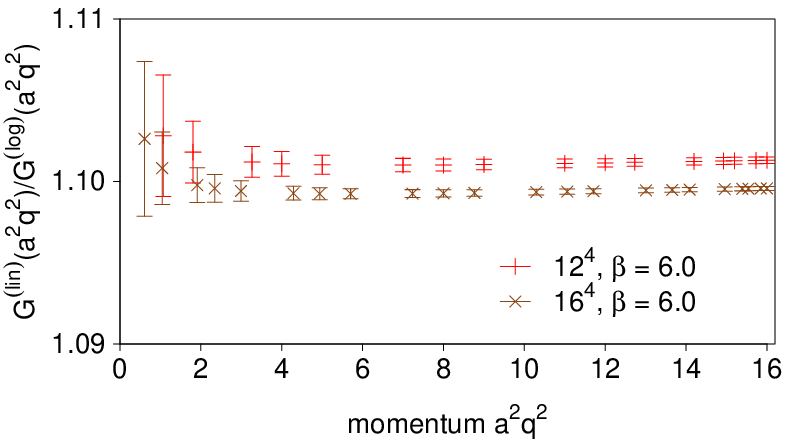} 
   \qquad 
   \includegraphics[angle=0]{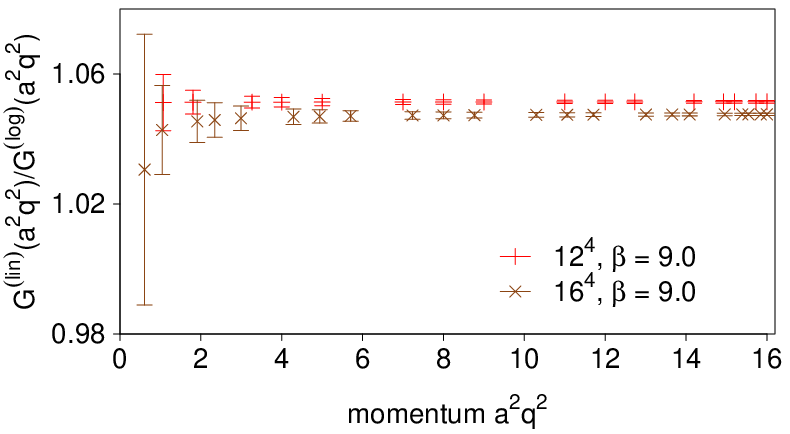}  }
\caption{The ratio $C_{\rm ghost}$ acc.\ to (\ref{eq:ghost_renorm_const}) 
  relating the ghost propagator for the two definitions of the gluon field 
  and the Faddeev-Popov operator, respectively, for $12^4$ and $16^4$ 
  lattices generated at $\beta = 6.0$ (left) and $\beta = 9.0$ (right).}
   \label{fig:figure10}
\end{figure*}

Similar to \Fig{fig:figure8}, in \Fig{fig:figure10} we show the ratio
\beq
  C_{\rm ghost}(q^2)=\frac{G^{\rm (lin)}(q^2)}{G^{\rm (log)}(q^2} 
  \label{eq:ghost_renorm_const}
\eeq
as a function of the momentum squared $q^2$.
For both values of $\beta$, we see an approximately constant ratio over 
a wide momentum range. The deviation seen at the smallest momenta 
for $\beta=9.0$ remains within statistical errors. 
In \Tab{tab:table5} we list the values for $C_{\rm gluon}$ and 
$C_{\rm ghost}$. As expected, their ratios happen to be related as  
\beq
  C_{\rm gluon} \cdot C_{\rm ghost}^2 \approx 1 \,.
  \label{eq:ratio_condition}
\eeq
This implies that the ghost-gluon coupling
$\alpha^{\mathsf{MM}}_s(q^2)$ (see \Sec{sec:runningcoupling}) 
determined directly from the gluon and ghost dressing functions is the
same for the logarithmic or linear definition.

%
\begin{table}[tb]
\begin{center}
\caption{The ratios $C_{\rm gluon}$ and $C_{\rm ghost}$
  [Eqs.~\eqref{eq:gluon_renorm_const} and
  \eqref{eq:ghost_renorm_const}] for the linear and logarithmic
  definition.}
\label{tab:table5}
\begin{tabular}{c@{\qquad}c@{\qquad\quad}c@{\qquad}c} 
\hline\hline
lattice & $\beta$ & $C_{\rm gluon}$ & $C_{\rm ghost}$    \\
\hline
$12^4$  & $6.0$   & $0.82 \pm 0.02$  & $1.1013 \pm 0.0007$ \\
$12^4$  & $9.0$   & $0.91 \pm 0.03$  & $1.0510  \pm 0.0010$  \\
$16^4$  & $6.0$   & $0.82 \pm 0.03$  & $1.0996 \pm 0.0006$ \\
$16^4$  & $9.0$   & $0.91 \pm 0.04$  & $1.0460  \pm 0.0040$  \\
\hline
\end{tabular}
\end{center}
\end{table}


\subsection{Finite-volume and lattice discretization effects}
\label{sec:finvol_discreteff}

Next we analyze discretization and finite-volume effects. In this
section the discussion will be restricted to the propagators for the
logarithmic definition. Corresponding data for ghost-gluon coupling,
$\alpha^{\mathsf{MM}}_s(q^2)$, is then discussed in the next section.

When checking lattice discretization artifacts, we fix the physical volume
such that it approximately equals that of all data for different
$a(\beta)$. In contrast, finite-volume effects will be analyzed for a
fixed $\beta$ varying the lattice size.

\begin{figure*}
        \centering
        \mbox{
        \includegraphics[angle=0]{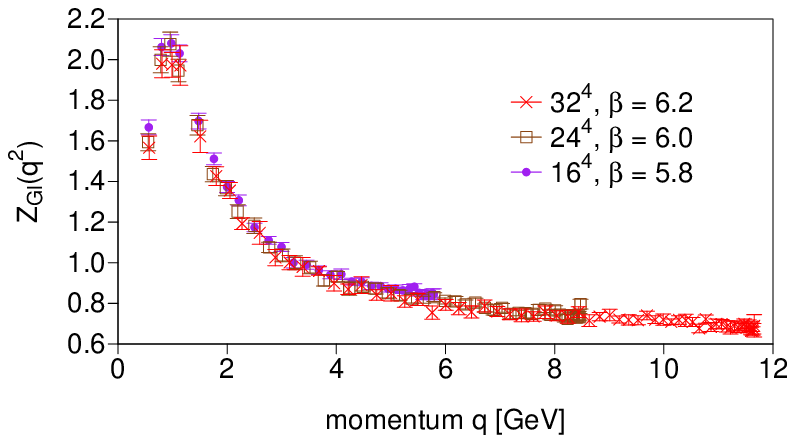}
        \qquad 
        \includegraphics[angle=0]{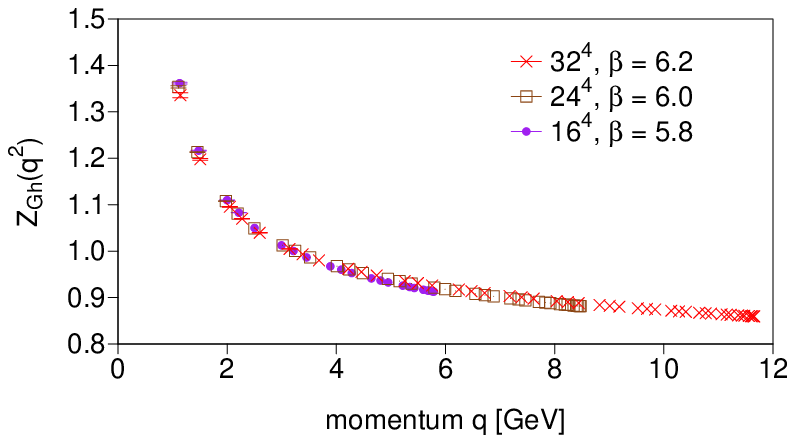}  }
        \caption{Renormalized gluon (left) and 
          ghost dressing function (right) for the logarithmic
          definition and various lattice spacings $a=a(\beta)$. The 
          physical volume is fixed to $V = (2.2 \mathrm{~fm})^4$. 
          Data has been renormalized at $q=\mu\approx 3.2\mathrm{~GeV}$.}
        \label{fig:figure11}
\end{figure*}

To analyze lattice discretization artifacts for the gluon and ghost
propagators we compare their renormalized dressing functions at
$\beta=5.8$, 6.0 and 6.2. Using the respective lattice sizes $16^4$,
$24^4$ and $32^4$, the physical volume is then roughly $V \simeq (2.2
\mathrm{~fm})^4$. For the renormalization we chose $\mu \approx 3.2
\mathrm{~GeV}$, which we find lies well below the momenta where
discretization artifacts could affect the renormalization. The
corresponding data is shown in \Fig{fig:figure11} suggesting 
that, with respect to precision of the data, lattice discretization
artifacts are reasonably small.

\begin{figure*}[tb]
        \centering
        \mbox{
        \includegraphics[angle=0]{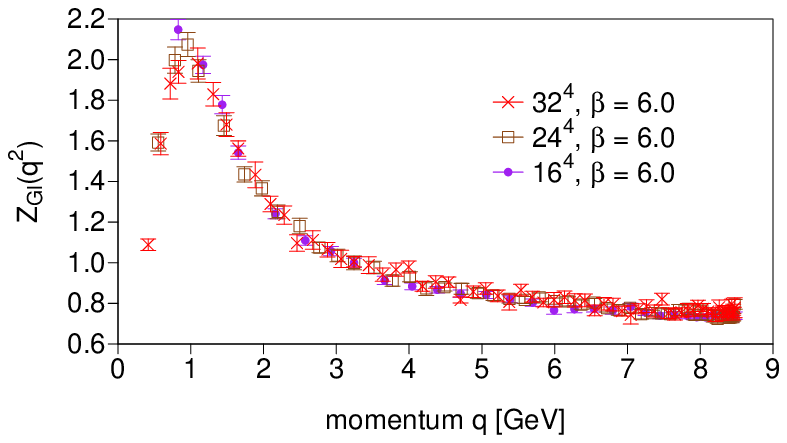}
        \qquad 
        \includegraphics[angle=0]{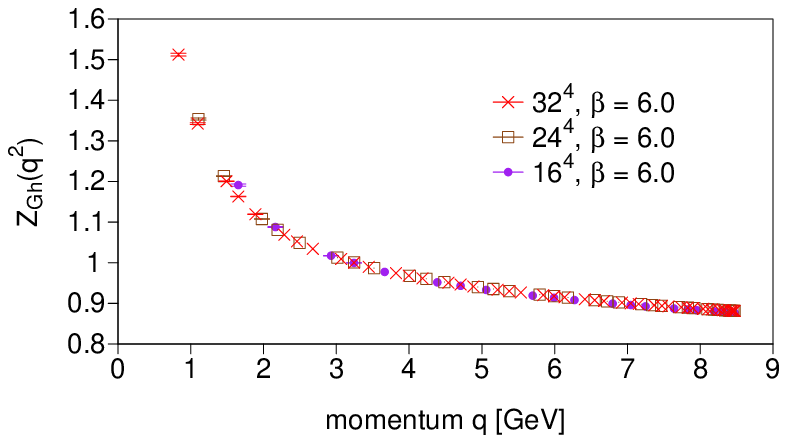}  }
        \caption{The renormalized gluon (left) and ghost dressing function 
        (right) for the logarithmic definition. The data is for a
        fixed lattice spacing ($\beta=6.0$) but three different
        volumes $V$. The renormalization point is at $\mu \approx  3.2
        \mathrm{~GeV}$.} 
        \label{fig:figure12}
\end{figure*}

To check finite-volume effects we choose $\beta=6.0$ and
vary the lattice size from $16^4$, $24^4$ and $32^4$. This has been
arranged for the data in \Fig{fig:figure12} where we compare the
renormalized gluon and ghost dressing functions. One clearly sees that 
finite-volume effects are negligible above $1 \mathrm{~GeV}$, the
momentum where the gluon dressing function has its maximum. At
${\cal O}(1) \mathrm{~GeV}$ and below a slight volume dependence
becomes visible. The overall behavior resembles that what has been observed
for $A^{\rm (lin)}$ in other studies. In order to see these effects well below 
$1 \mathrm{~GeV}$, much larger lattices are needed, for example, as
those studied for the linear definition in~\cite{Bogolubsky:2009dc}. 

\subsection{The running coupling}
\label{sec:runningcoupling}

The running coupling $\alpha_s(q^2)$ for Yang-Mills theories can be 
defined in various ways. Here we use the coupling of ghost-gluon
vertex in a particular (minimal) MOM scheme (see
\cite{vonSmekal:1997is,Alkofer:2000wg} as well as the more recent
papers \cite{vonSmekal:2009ae,Sternbeck:2010xu}). It can be defined in
terms of the bare, i.e., unrenormalized gluon and ghost dressing
functions $Z_{\rm Gl}$ and $Z_{\rm Gh}$ as follows:
\beq
  \alpha^{\mathsf{MM}}_s(q^2) = \frac{g_0^2}{4 \pi} Z_{\rm Gl}(a^2,q^2)Z_{\rm Gh}^2(a^2,q^2)\,. 
  \label{eq:running_coupling}
\eeq
It is a renormalization-group invariant quantity, i.e., shifting the 
cutoff $a^{-1}$ or transforming the right hand side into renormalized 
quantities and changing their subtraction momentum $\mu$ within the given 
MOM scheme should not alter $\alpha^{\mathsf{MM}}_s(q^2)$.
Therefore, we can compute it directly from the bare lattice dressing
functions at an arbitrary large enough cutoff-value $a^{-1}(\beta)$,
as long as multiplicative renormalization is ensured and additive
lattice artifacts are suppressed. In what follows we shall omit the 
superscript $\mathsf{MM}$ for simplicity.

\begin{figure*}[tb]
   \centering
   \mbox{
   \includegraphics[angle=0]{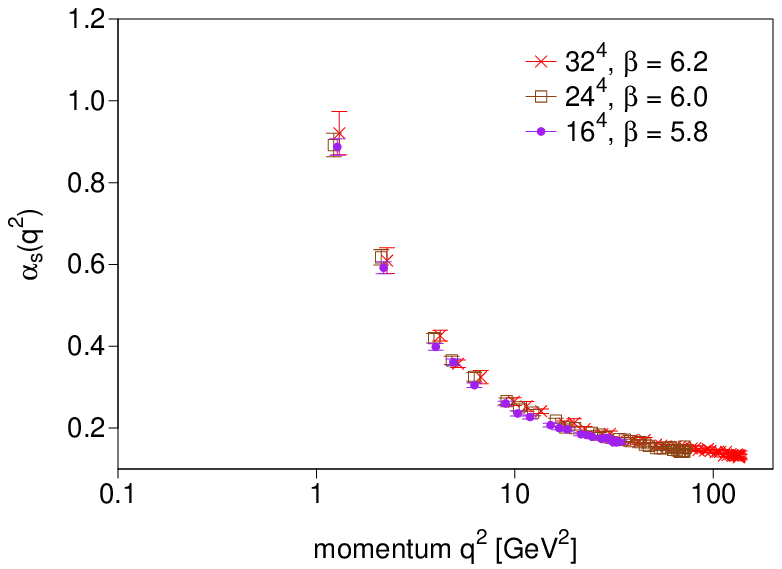}
   \qquad 
   \includegraphics[angle=0]{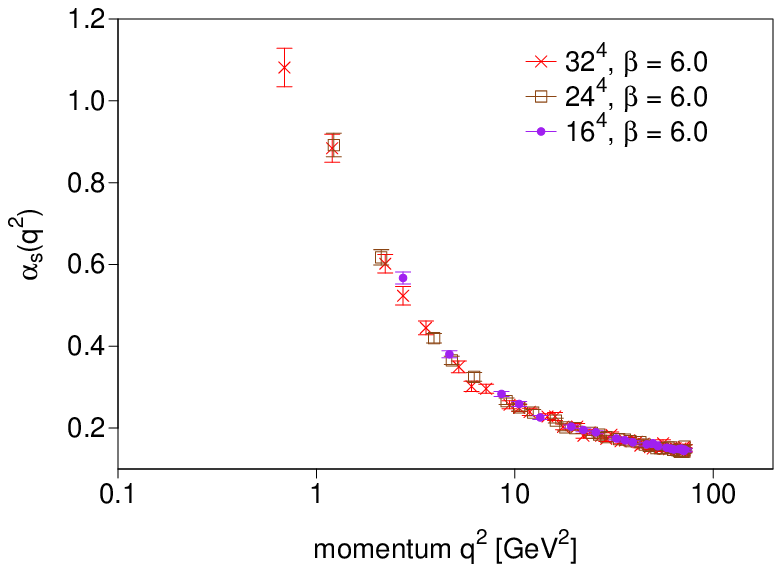}  }
   \caption{$\alpha_s(q^2)$ for the
     logarithmic definition as a function of physical momentum
     squared. Left: for three lattice spacings $a=a(\beta)$ but
     fixed physical volume $V = (2.2 \mathrm{~fm})^4$. Right: for
     different volumes $V$ but fixed lattice spacing $a=a(\beta=6.0)$.} 
    \label{fig:figure13}
\end{figure*}

First, we check effects due to the lattice discretization and the
finite volume. In order to investigate lattice discretization effects
we present on the left hand side of \Fig{fig:figure13} the running
coupling $\alpha_s(q^2)$ 
for different lattice spacings but fixed physical volume [as above we 
choose again $V = (2.2 \mathrm{~fm})^4$]. Apparently, there are some 
systematic lattice discretization effects, suggesting that for the given 
(rather small) $\beta$ additive lattice artifacts are small but not
negligible. This is in agreement with the findings in
\cite{Sternbeck:2010xu}. These artifacts should disappear for large 
$\beta$.
On the right hand side of \Fig{fig:figure13}, we show data for different
physical lattices sizes but fixed lattice spacing [again we choose
$\beta=6.0$]. Based on that figure we have to conclude that finite-volume
effects seem to be negligible for the considered momentum range.

\begin{figure}[tb]
    \centering
    \includegraphics[angle=0]{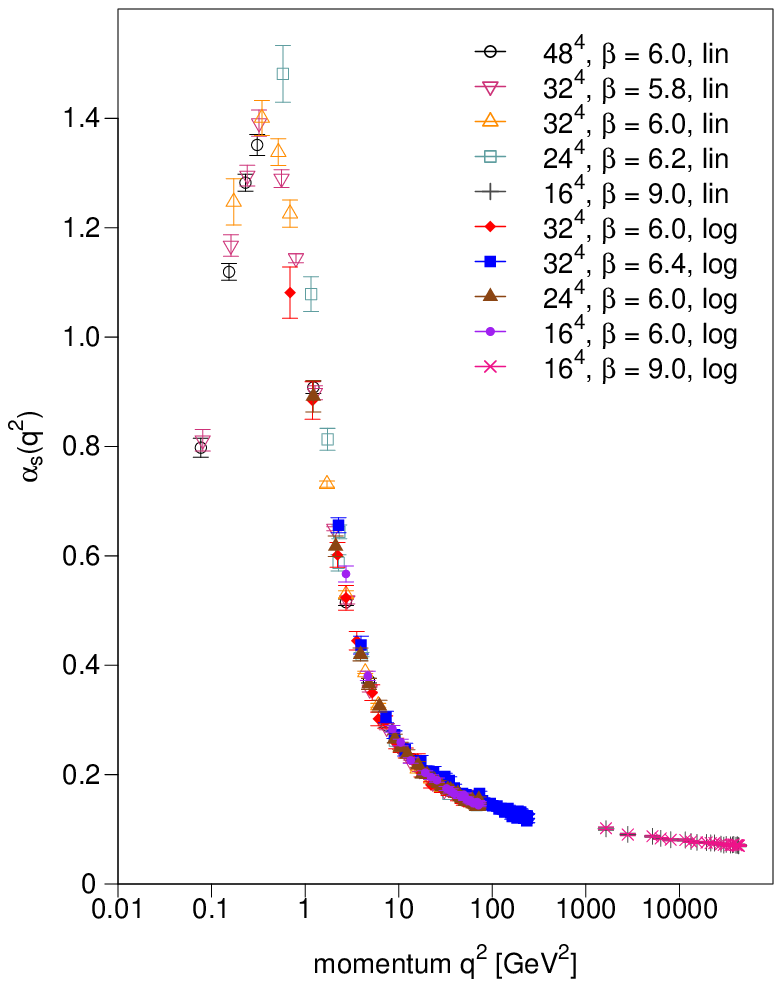}
\caption{Running coupling for various lattice sizes and
  $\beta$-values. Filled symbols are for the logarithmic definition,
  while open symbols are for the linear one.}
    \label{fig:figure14}
\end{figure}

In \Fig{fig:figure14} we finally compare the coupling for the
logarithmic and the linear definition. We decided to show data for various
lattice sizes and $\beta$ values in a single plot to demonstrate that
altogether the data for the two definitions matches up almost
completely. Regrettably, there are some small deviations due to the
different lattice spacings and volumes, but this should be
expected. The almost perfect overlap of the two curves agrees, of
course, well with what we saw above for the ratios of the propagators
(see \Eq{eq:ratio_condition}). That is, $\alpha_s(q^2)$ will not
differ calculated either for the standard (linear) approach (as in
\cite{Sternbeck:2010xu}) or for the logarithmic approach as done
here. Unfortunately, we cannot say which approach comes with the
smaller lattice discretization artifacts. This is left for a future study. 

\section{Comparison with NSPT} 
\label{sec:compareresultslogversusNSPT}

We now turn to the NSPT results 
of~\cite{DiRenzo:2009ni,DiRenzo:2010cs,DiRenzo:2010ep} which below
we will compare to our data. Let us start first with some
facts on NSPT.

NSPT is a numerical approach to lattice perturbation theory that allows
to circumvent the difficulties of the standard diagrammatic
approach and facilitates automatized perturbative
calculations. It has its roots in stochastic quantization and is based
on a modified Langevin equation equipped with stochastic gauge fixing
corresponding to a gauge fixing term
$(\partial_\mu A_\mu)^2/(2\xi)$ at finite $\xi$~\footnote{For a study 
of an {\it approximate} Landau gauge within a Langevin approach see
Ref.~\cite{Pawlowski:2009iv}.}. 
For our purposes it is actually
a hierarchy of first-order evolution equations associated with the
various parts of the gauge field when expanded in
terms of the lattice coupling $g_0 \propto 1/\sqrt{\beta}$:
\begin{align}
 U_{x,\mu} &= 1+ \sum_{l \ge 1} \beta^{-l/2} U^{(l)}_{x,\mu} \,, \\
 A^{\rm (log)}_{x+\frac{\hat{\mu}}{2},\mu} &=
    \sum_{l \ge 1} \beta^{-l/2} A^{(l)}_{x+\frac{\hat{\mu}}{2},\mu} \,, \nonumber
\end{align}
From a numerical point of view, these different parts, representing
first, second, third etc.\ orders, are separately dealt within the
code. The maximal addressable order of perturbation theory
is thus limited by the available computing resources (cpu time and memory). 

The Langevin simulation is implemented in an Euler scheme with a
finite evolution time step. Lattice observables, in our case the
ghost~\cite{DiRenzo:2009ni} and gluon~\cite{DiRenzo:2010cs}
propagators, are evaluated taking the long-time average, order by order
in a loop expansion in even powers of $g_0$. Contributions from odd
powers vanish within the statistical errors.  As for any Langevin
simulation, one then has to take the limit to vanishing time
step. This is in addition to the continuum limit and the limit of
infinite volume.

Regardless of this, NSPT results for a {\it finite} lattice volume can  
be confronted directly with standard MC results for a given $\beta$,
supposed the lattice size and definition of the studied
observable is the same in both approaches.

But before such a comparison is possible, the limit $\xi\to 0$ 
(\emph{minimal Landau gauge}) has to be taken. This is arranged such
that a sequence of configurations (separated by ${\cal O}(50)$
Langevin time steps) undergoes a Fourier-accelerated gauge-fixing
procedure, after which the individual gluon fields, $A^{(l)}_\mu$,
each associated with particular perturbative order ($g_0^l$), are
transversal within machine precision.

As in standard lattice perturbation theory, an expanded version of the
logarithmic relation between the gluon fields and the transporters
(compare (\ref{eq:Alog})) is taken into account up to the maximal order
of perturbation theory addressed in the given case. 
Correspondingly, the gauge
functional and the structure of the Faddeev-Popov operator are the same
as in Eqs.~\eqref{eq:EQ33} and \eqref{eq:EQ34}.

The gluon two-point function in $n$-loop order is then defined as a 
convolution of the bilinears of gluon fields (in momentum space) in
complementary orders:
\begin{equation}
  \delta^{ab} D_{\mu\nu}^{(n)}(p(k)) = \left\langle \,
  \sum_{l=1}^{2n+1}
  \left[ \widetilde{A}^{a,(l)}_{\mu}(k) \,
  \widetilde{A}^{b,(2n+2-l)}_{\nu}(-k) \right] \,
  \right\rangle_U \,.
\label{eq:Dn}
\end{equation}
The Faddeev-Popov operator (explicitly written in \Eq{eq:EQ34} up 
to fourth order) can be expanded in terms of products of various $A^{(l)}$, 
with the term $M^{(n)}$ collecting all terms of order $g_0^n$.
This structure allows to express the inverse of the Faddeev-Popov 
operator also as an expansion in orders of $g_0$ in a recursive way, 
without the need of explicitly inverting any other than the zeroth
order term, \hbox{$M^{(0)} = \Delta$} (the Laplacian).

A reasonable ``convergence'' of the NSPT results up to few loops
(three or four are available now) requires a small bare coupling $g_0$,
i.e., a large $\beta$.  However, the bare coupling $g_0$ is known to be
a poor expansion parameter~\cite{Lepage:1992xa}.  One can speed up
convergence by ``boosting'', i.e., trading the bare coupling constant
by an effective ``boosted'' coupling $g_b^2 =
g_0^2/P_{\mathrm{pert}}(g_0^2) > g_0^2$, where $P_{\mathrm{pert}}$ is defined by the
average perturbative plaquette. Its expansion is
determined within the Langevin simulations, along with the
propagators. The effect of the larger boosted coupling is
overcompensated by the rapid decay of the expansion coefficients with
respect to $g_b^2$.
\begin{figure*}
   \centering
   \mbox{
   \includegraphics[width=8.0cm,angle=0]{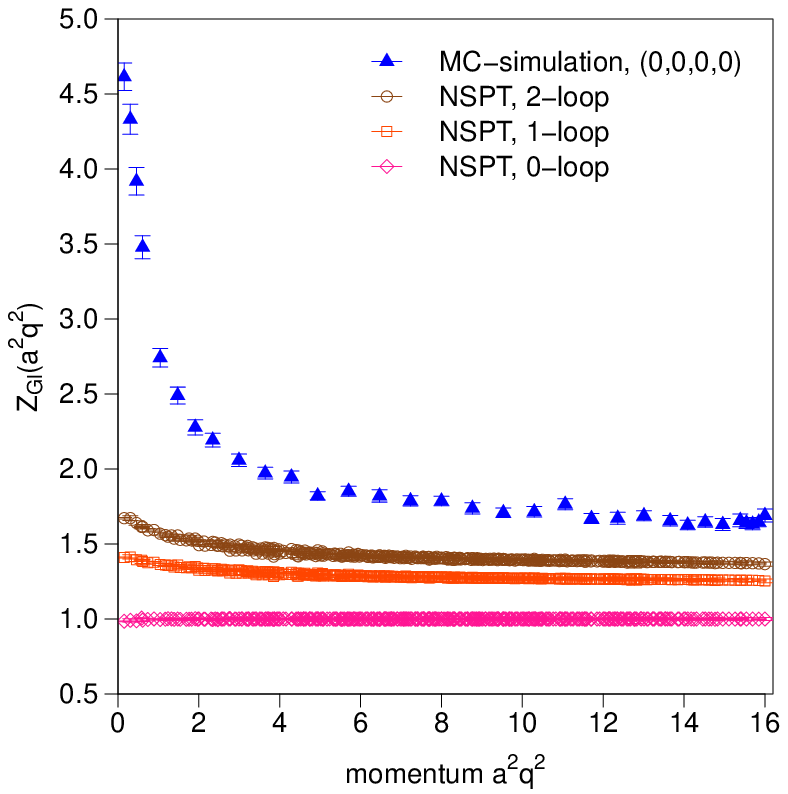}
   \qquad
   \includegraphics[width=8.0cm,angle=0]{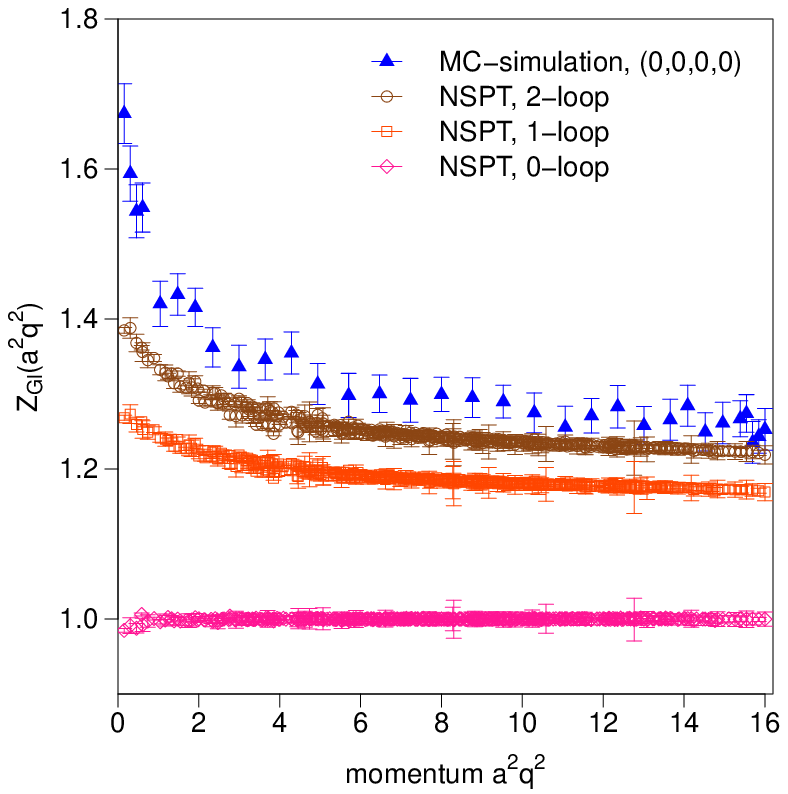}  }
\caption{Comparison of MC with NSPT results 
  for the bare gluon dressing function at (cumulative) tree, 1-loop 
  and 2-loop level at $\beta=6.0$ (left) and $\beta=9.0$ (right). 
  The lattice size is $16^4$.}
\label{fig:figure15}
\end{figure*}

\begin{figure*}
   \centering
   \mbox{
   \includegraphics[width=8.0cm,angle=0]{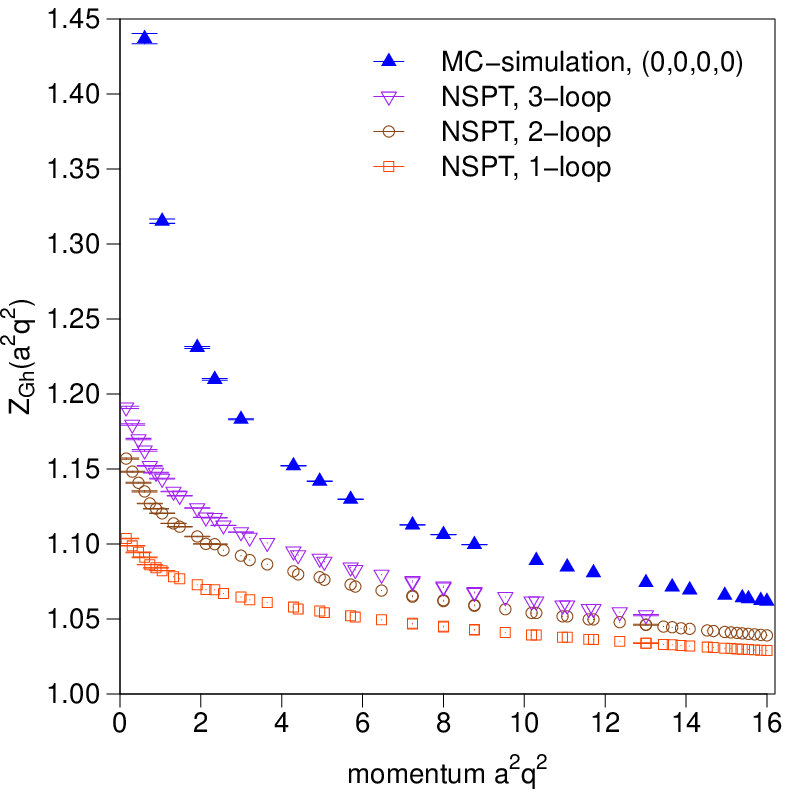}
   \qquad
   \includegraphics[width=8.0cm,angle=0]{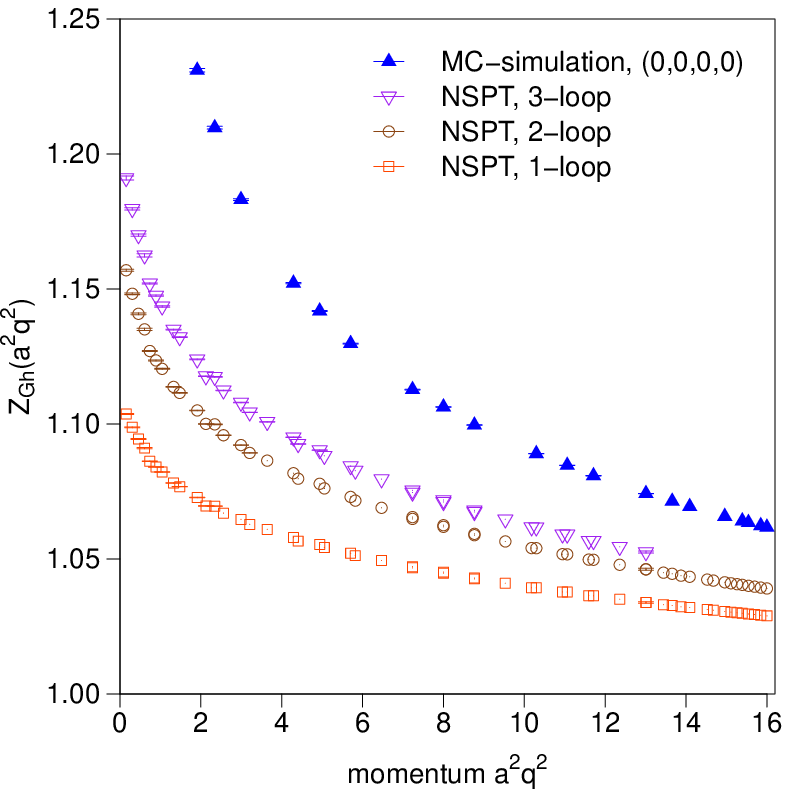}  }
\caption{Comparison of MC with NSPT results 
for the bare ghost dressing function at (cumulative) 1-loop, 2-loop 
and 3-loop level at $\beta=6.0$ (left) and $\beta=9.0$ (right). 
The lattice size is $16^4$.}
   \label{fig:figure16}
\end{figure*}

\begin{figure*}[tb]
   \centering
   \mbox{
   \includegraphics[width=7.6cm,angle=0]{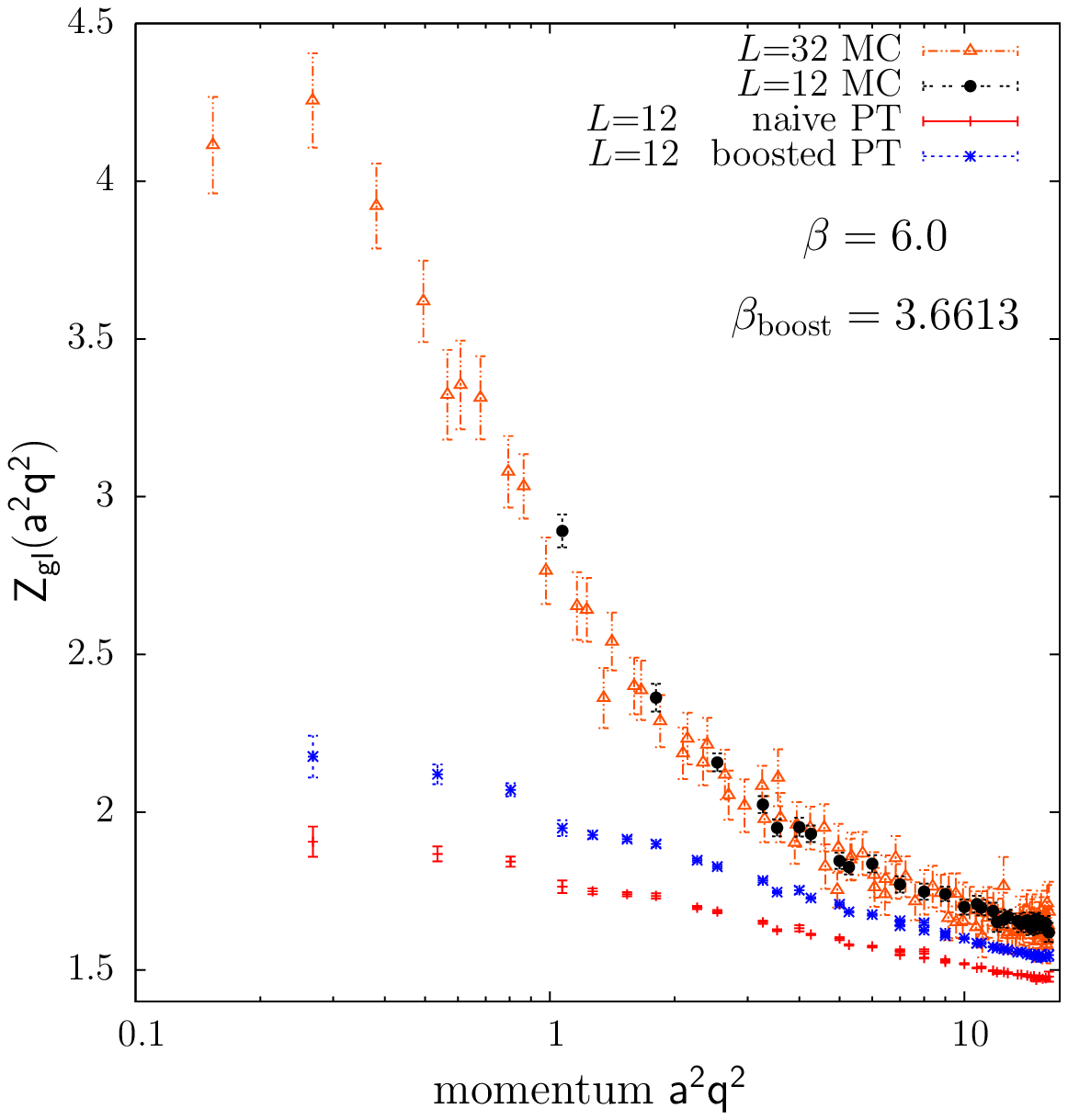}
   \qquad
   \includegraphics[width=7.6cm,angle=0]{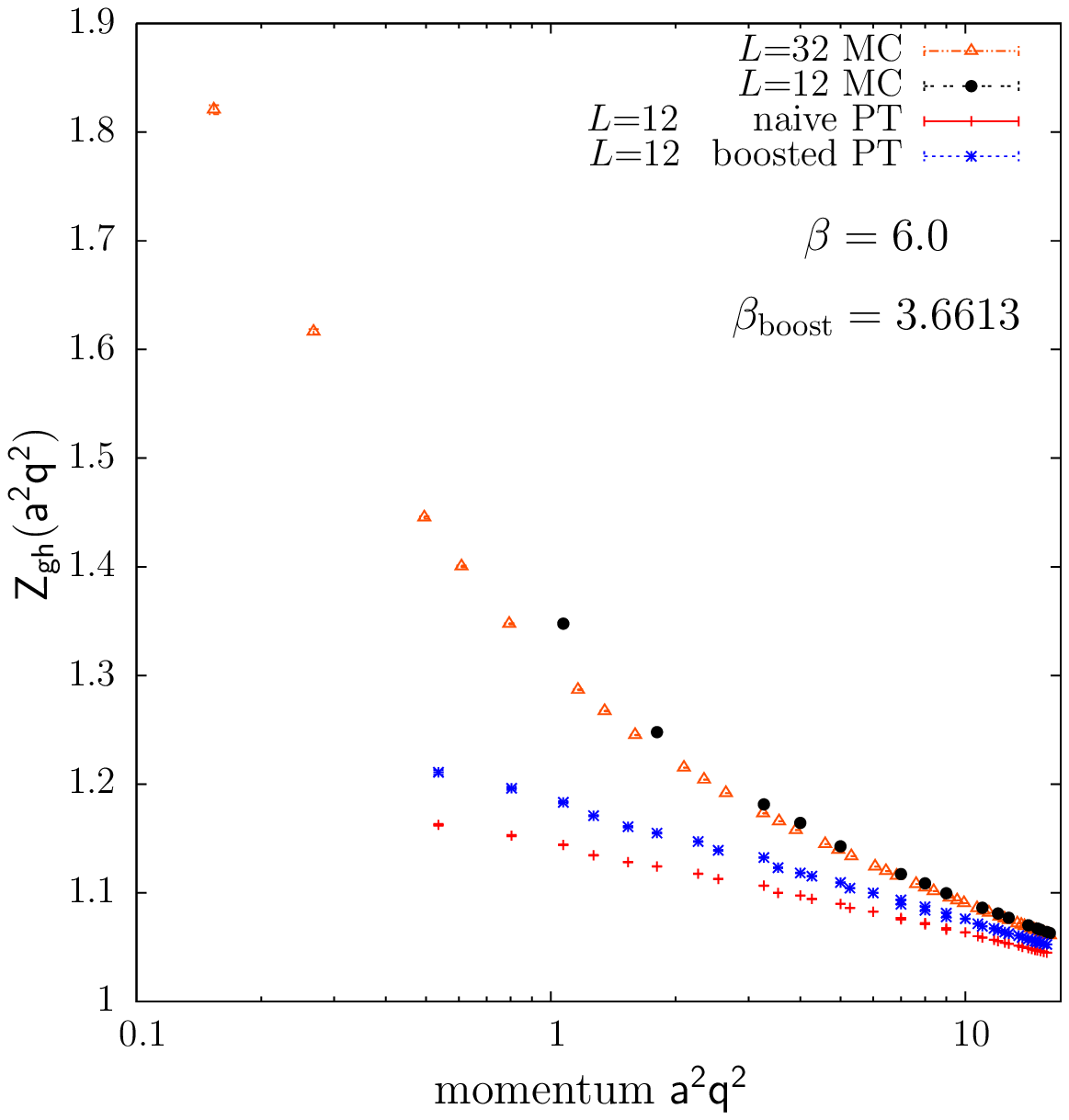}  }
\caption{Comparison of naive and boosted LPT (based on NSPT) with MC data 
  for a $12^4$ lattice at $\beta=6.0$. 
  Left: the bare gluon dressing function up to 4-loop.
  Right: the bare ghost dressing function up to 3-loop.
  We also include MC data for a $32^4$ lattice.}
\label{fig:figure17}
\end{figure*}

\begin{figure*}[tb]
   \centering
   \mbox{
   \includegraphics[width=8.0cm,angle=0]{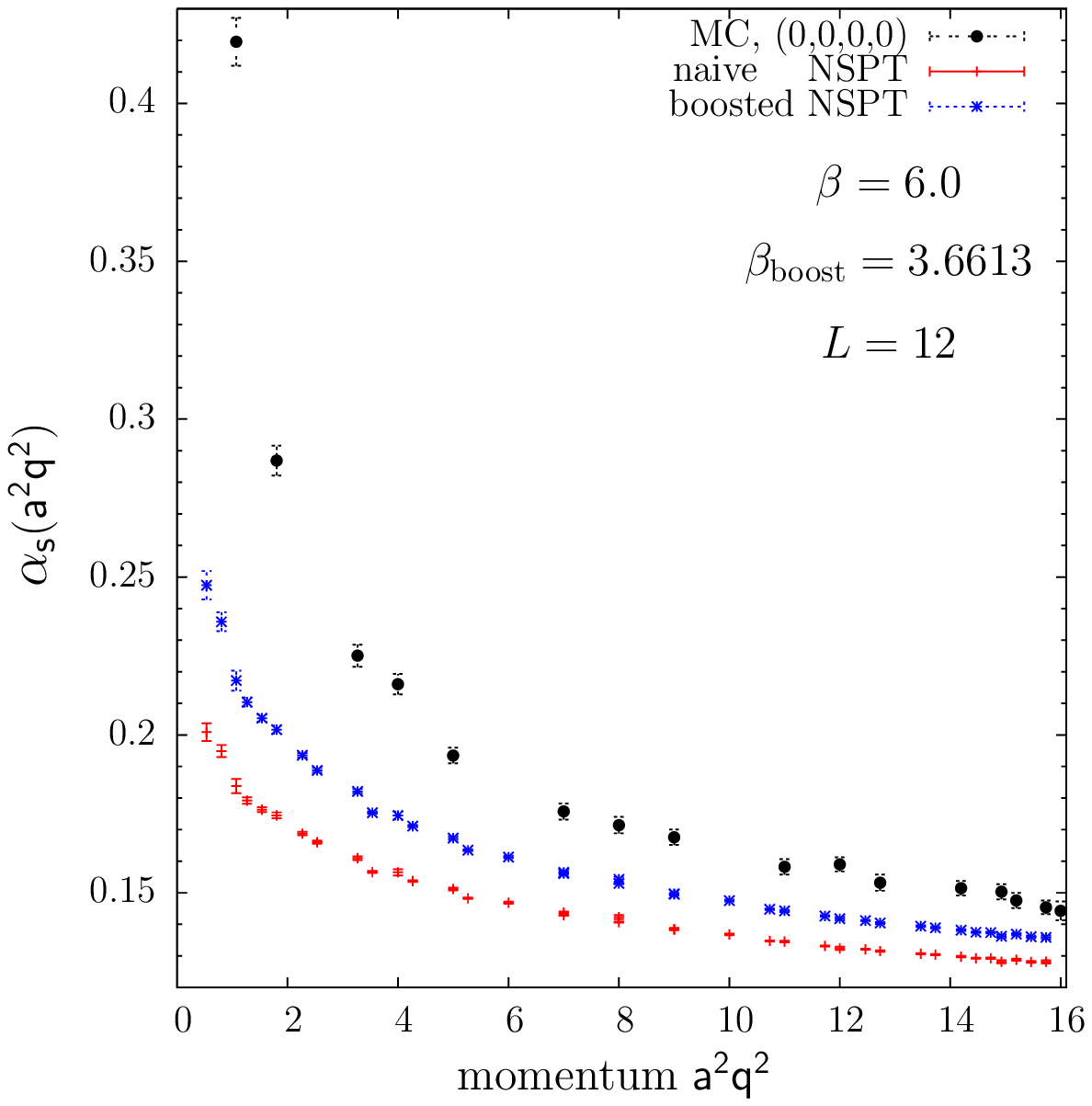}
   \qquad
   \includegraphics[width=8.0cm,angle=0]{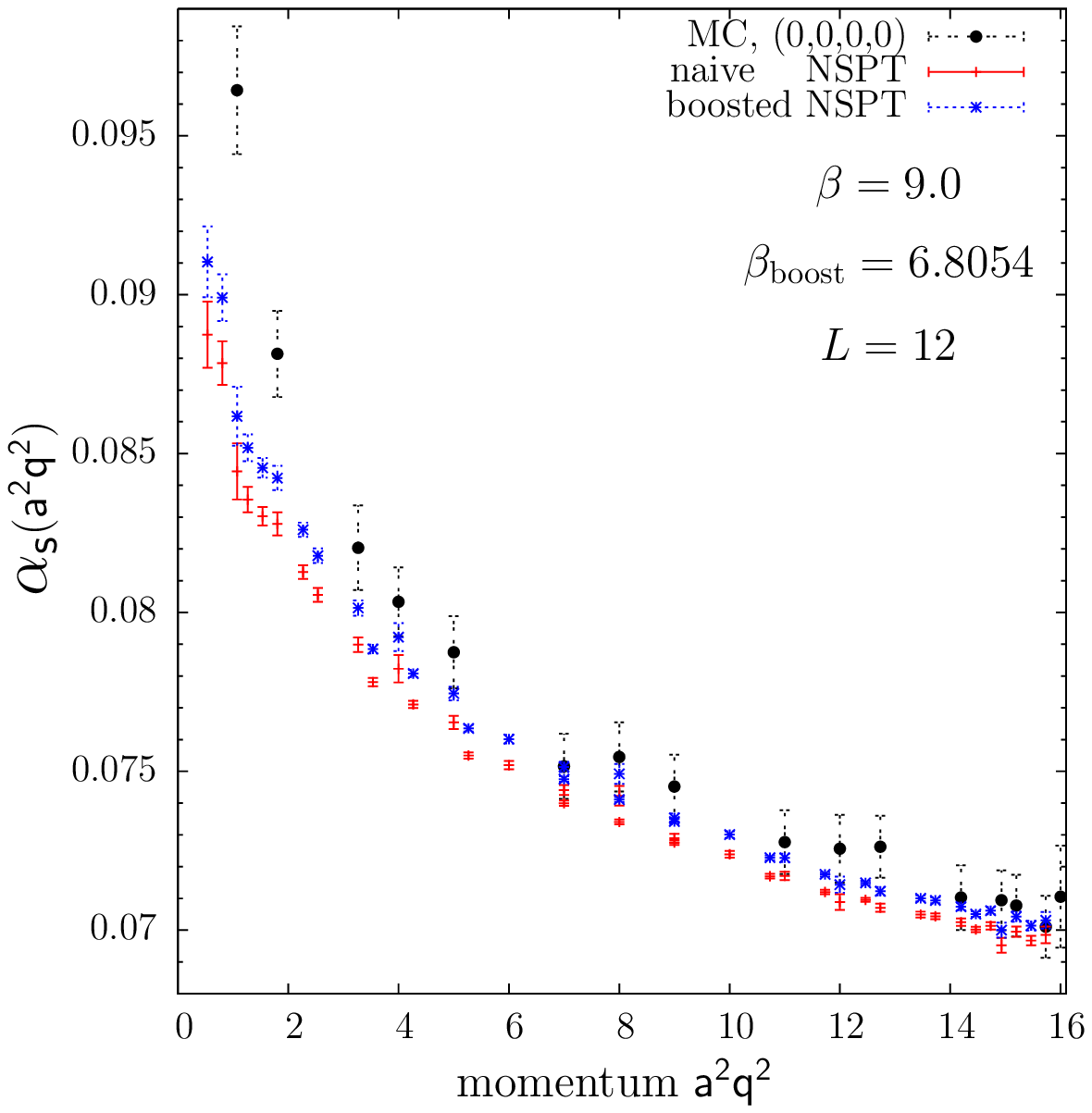}  }
 \caption{Comparing naive and boosted LPT (based on NSPT) data
   for the running coupling constant $\alpha_s(q^2)$ 
   to corresponding MC data for a $12^4$ lattice. Left: $\beta=6.0$.
   Right: $\beta=9.0$. For the LPT data, the gluon (ghost) dressing
   function up to 4-loop (3-loop) accuracy has been included.}
\label{fig:figure18}
\end{figure*}

We now compare our data for the logarithmic definition to the NSPT
results of~\cite{DiRenzo:2009ni,DiRenzo:2010cs,DiRenzo:2010ep}. 

One should be aware that on small lattices, with sizes like $12^4$ or 
$16^4$, and for the larger $\beta$ values, the Monte Carlo lattice gauge
fields will be in a pseudo-deconfinement phase.
This can be monitored by the ``spatially'' averaged Polyakov loops,
actually in all four directions. It is well-known that the agreement 
of Monte Carlo results with standard LPT requires that the Monte Carlo
simulation is guaranteed to stay in the trivial Polyakov 
sector~\cite{Damm:1998pd}. For $SU(3)$ this means, that the average 
Polyakov loops in all four directions have to be located in the sector 
of predominantly real values. Let us denote averages in this sector by 
the corresponding phases $(0,0,0,0)$, in distinction to results 
obtained from Monte Carlo configurations without taking notice of the
Polyakov sector.
Note that one can easily switch between the Polyakov loop sectors by
applying global $Z(3)$ transformations on all link variables attached
to and pointing forward, orthogonal to an arbitrary 3D plane.  While
at $\beta=9.0$ the MC results for the gluon and ghost dressing
functions clearly depend on the Polyakov loop sector, for
$\beta=6.0$ the Polyakov loop values fluctuate closely around the
origin of the complex plane, i.e., in the confined phase. In this case
the choice of a sector should not influence the behavior of the
propagators or dressing functions.  In what follows, for the detailed
comparison all Monte Carlo configurations at $\beta=9.0$ have been
flipped to the real sector $(0,0,0,0)$ if necessary, before the gauge
fixing has been performed.

Let us first confront the tree level and the cumulative one-loop and
two-loop contributions to the gluon dressing function with the results
from Monte Carlo simulations (see \Fig{fig:figure15}). The simulations
have been performed for a $16^4$ lattice at $\beta=6.0$ and
$\beta=9.0$, the same values as for the NSPT results. For the reader's
convenience, we present the NSPT data the same way the Monte Carlo
data has been present above. When looking at the data in \Fig{fig:figure15}
we see the NSPT results approaches the MC data with increasing loop
order. As expected though, this is less the case for $\beta=6.0$, but for
$\beta=9.0$ the NSPT data almost approaches the MC data.

Similar we see for the bare ghost dressing function in
\Fig{fig:figure16}. For this the speed of convergence of the
NSPT results can be assessed from the difference between 
two-loop and three-loop~\cite{DiRenzo:2009ni}. The three-loop result
is already very close to the MC result for $\beta=9.0$ and at the
largest available momenta. 

Next we illustrate the effect of ``boosting" the perturbative
expansion. For this we use the currently available NSPT results
(up to four/two loops ($L\le 12$/$L=16, 20, 32$) for the gluon
propagator and up three loops ($L\le 20$) for the ghost propagator)
and confront them with corresponding MC data. This is shown in
Fig.~\ref{fig:figure17} where also the bare inverse coupling $\beta$
and its boosted value $\beta_{\mathrm{boost}}$ are given. As expected,
boosting moves the NSPT data closer to the MC results, but they cannot
be reached, certainly not at $\beta=6.0$.  

Last but not least, the running coupling, $\alpha_s(q^2)$, 
as calculated from the NSPT dressing functions, both summed up to the
available orders, is compared to the Monte Carlo results at
$\beta=6.0$ and $9.0$. The corresponding data is shown in
Fig.~\ref{fig:figure18}, again for naive and boosted perturbation
theory. We see that the running coupling from Monte Carlo simulations
is approached from below up to 7\% for $\beta=6.0$ and practically
approached within the present errors for $\beta=9.0$.

Concluding this section, one can say that our MC results 
for the logarithmic approach to Landau gauge propagators 
support the validity of the NSPT calculation for the gluon as well as 
for the ghost propagator. In fact, the NSPT results do not coincide
with the MC data but they become closer with increasing order of the 
perturbative expansion. Also, the difference between the NSPT and MC data
becomes smaller for larger $\beta$, suggesting the difference to be 
related to nonperturbative effects NSPT results cannot provide.


\section{Conclusions}
\label{sec:conclusions} 

In this paper we have studied an alternative approach to compute the
$SU(3)$ Landau gauge gluon and ghost propagators on the lattice. This
approach uses a logarithmic ansatz [\Eq{eq:Alog}] for the definition
of the lattice gluon fields from a given gauge field configuration. It
is thus best suited to compare lattice MC data for these propagators
with results from NSPT. 
We have started the task by first exploring some options for an efficient
algorithm that fixes gauge field configurations to Landau gauge for
the logarithmic case. We find a multigrid-accelerated gradient method
with a preconditioning step of simulated annealing and subsequent
overrelaxation, applied to the linearly defined gluon field, to
be a good choice. The method is also easy parallelizable. 

With this algorithm at our disposal we have compared the bare lattice
propagators for the logarithmic and the linear definition of gluon
fields. As expected, we see them to differ by multiplicative factors
that depend on $\beta$. Those factors are such that they perfectly
cancel when one considers $\alpha_s(q^2)$. Apart from some 
small lattice discretization artifacts, data for the running coupling
matches up almost completely, as it should for a renormalization-group
invariant object. It is thus also an ideal quantity to assess
discretization artifacts.

For the logarithmic definition we have checked Gribov copy, finite-size and
lattice discretization effects, and find them to be small for momenta
$q > 1 \mathrm{~GeV}$.

Finally we have compared our MC data for the logarithmic definition
with results from NSPT. These are available up to four loops for the
gluon propagator, and up to three loops for the ghost propagator. We
find a reasonable convergence at large momentum. Note that for this it
is important that during the MC process the gauge field configurations are
kept in the correct (real-valued) Polyakov loop sector. For large $\beta$,
these may easily pass a pseudo-deconfinement phase transition.

Our results altogether support universality with respect to the two
lattice realizations of $SU(3)$ Landau gauge theory studied
herein. In as far this universality persist in the low-momentum region
remains to be seen.

We emphasize that the universality of different lattice definitions 
we have observed in this paper assumes a unique Landau gauge fixing 
based on the (global) minimization of a corresponding gauge functional. 
For an alternative view of dealing with the Gribov copy problem
see Refs. \cite{Maas:2009se,Maas:2009ph}.  

\section*{Acknowledgments}

The authors are grateful for the generous support by the HLRN Berlin/Hannover
with CPU time on their massively parallel supercomputing system. 
E.-M.I. and M.M.-P. acknowledge financial support by the DFG via
Mu932/6-1. A. St. acknowledges support from the European Union through
the FP7-PEOPLE-2009-RG program and by the SFB/Tr~55.


\appendix
\section{Multigrid Fourier-accelerated gauge-fixing}
\label{sec:appendix} 

For our implementation of a parallel version of the Fourier-accelerated
gauge fixing we follow Goodman and Sokal
\cite{Goodman:1986pv,Goodman:1989jw} using the representation of the
Fourier transformation of $1/q^2$ in position space:
\bea
&&  \hat{F}^{-1}\left[\frac{q_{\rm max}^2}{q^2} 
    \hat{F}\left[(\sum_\mu\partial_\mu {}^g \! A^{\rm (log)}_\mu)(x)\right]\right]  \nonumber \\
&=& q_{\rm max}^2~\Delta^{-1} (\sum_\mu \partial_\mu {}^g \! A^{\rm (log)}_\mu)(x) \,.
  \label{eq:multigrid1}
\eea

This leads us to a simple inversion of the Laplacian, i.e., to solving the 4D 
Poisson equation
\beq
- \left( \Delta v \right)(x) = \left( \sum_\mu\partial_\mu A^{\rm (log)}_\mu \right)(x)
\label{eq:poisson}
\eeq
using for example the local Jacobi method. To avoid
critical slowing down inherent to this method we use the multigrid 
algorithm by solving (\ref{eq:poisson}) successively on the original 
(fine) lattice and on several coarser lattices. In order to use our multigrid 
algorithm for various lattice sizes we implement the multigrid with a symmetric 
lattice decomposition.

\begin{figure}[tb] 
  \centering
  \includegraphics{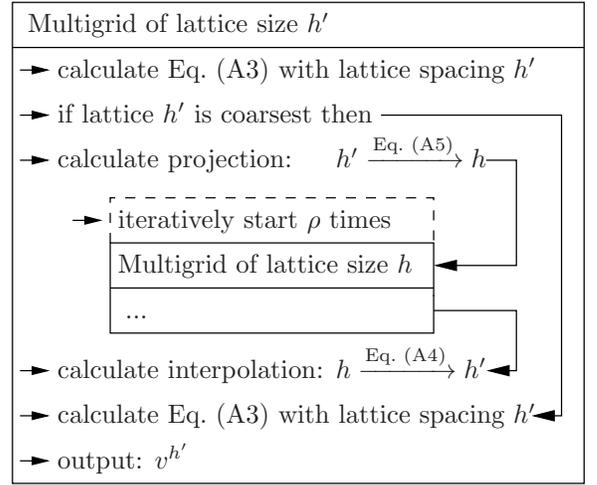}
  \caption{Flow chart illustrating our multigrid algorithm.}
  \label{fig:flowchart}
\end{figure}

On each sublattice one has to solve a system of linear equations
\beq
\label{eq:mg_main}
A^h v^h = f^h\,,
\eeq
where the superscript $h$ describes the respective lattice spacing 
$h = a, 2a, 4a, 8a, ..$. Comparing with (\ref{eq:poisson}) we can see that 
$A^a = \Delta$. For switching between the lattices one defines 
interpolation matrices $I$ as
\beq
\label{eq:mg_I}
\begin{split}
& \hspace{0.6cm} A^{h^{\prime}} = 
I_{h^{\prime},h} A^h \left[I_{h^{\prime},h}\right]^T \\
& v^{h^{\prime}} = I_{h^{\prime},h} v^h \hspace{0.2cm} , 
\hspace{0.2cm} f^{h^{\prime}} = I_{h^{\prime},h} f^h
\end{split}
\eeq
and projection matrices $P$
\beq
\label{eq:mg_P}
\begin{split}
& \hspace{0.6cm} A^h = P_{h,h^{\prime}} A^{h^{\prime}} 
\left[P_{h,h^{\prime}}\right]^T \\
& v^h = P_{h,h^{\prime}} v^{h^{\prime}} \hspace{0.2cm} , 
\hspace{0.2cm} f^h = P_{h,h^{\prime}} f^{h^{\prime}} \,.
\end{split}
\eeq
$h^{\prime}$ always denotes the lattice with the finer spacing while 
$h$ denotes the coarser lattice. The matrix structure (in terms of the lattice 
sites) of these equations is left implicit.  The projection matrices are the 
transposed interpolation matrices (with respect to the indices pointing to 
lattice sites)
\beq
P_{h,h^{\prime}} = \left[ I_{h^{\prime},h} \right]^T \,.
\eeq
The matrices were chosen in the way that the operator $A^{h^{\prime}}$ 
got the same structure on all sublattices 
(i.e. $A^{h^{\prime}} = \Delta^{h^{\prime}}$).
In practice, the multigrid algorithm is realized by jumping between the finest 
and various coarser lattices. This is summarized in the flow chart in
\Fig{fig:flowchart}.

Solving \Eq{eq:mg_main} before the projection is called pre-smoothing, 
post-smoothing after the interpolation, respectively~\cite{Briggs:2000xx}. 
To solve \Eq{eq:mg_main} we use the Jacobi method with a fixed number of 
$20$ iterations. Hence we did not solve equation \Eq{eq:mg_main} at high 
numerical accuracy on the sublattices. Nevertheless, the accuracy of this 
calculation seems to have only a minor influence on the total number of 
iterations needed to fix the Landau gauge. For the parameter $\rho$ we 
chose the value $\rho = 2$, i.e., the so called $W$-cycle. 

\newpage

\vspace{-1.5cm}

\end{document}